\documentclass[twocolumn,trackchanges]{aastex63}
\usepackage{amsmath}
\usepackage{float}
\usepackage{url}
\usepackage{graphicx}
\usepackage{threeparttable} 
\usepackage{multirow}
\usepackage[figuresright]{rotating}
\usepackage{lipsum}
\usepackage{CJKutf8}
\usepackage{enumitem}
\usepackage{cleveref}
\usepackage{booktabs}
\usepackage{microtype}
\usepackage{tabularx}
\usepackage[normalem]{ulem}

\shorttitle{M-Type Stars Catalog}
\shortauthors{Li et al.}

\graphicspath{{./}{../}}

\begin{document}
\begin{CJK*}{UTF8}{gbsn}
\title{Refined M-type Star Catalog from LAMOST DR10: Measurements of Radial Velocities, $T_\text{eff}$, log $g$, [M/H] and [$\alpha$/M]}

\correspondingauthor{Yin-Bi Li $\&$ A-Li Luo}
\email{* ybli@bao.ac.cn lal@nao.cas.cn   }

\author[0000-0002-8913-3605]{Shuo Li}
\affiliation{CAS Key Laboratory of Optical Astronomy, National Astronomical Observatories, Beijing 100101, China}
\affiliation{University of Chinese Academy of Sciences, Beijing 100049, China}

\author[0000-0001-7607-2666]{Yin-Bi Li$^{*}$}
\affiliation{CAS Key Laboratory of Optical Astronomy, National Astronomical Observatories, Beijing 100101, China}
\affiliation{University of Chinese Academy of Sciences, Beijing 100049, China}

\author[0000-0001-7865-2648]{A-Li Luo$^{*}$}
\affiliation{CAS Key Laboratory of Optical Astronomy, National Astronomical Observatories, Beijing 100101, China}
\affiliation{University of Chinese Academy of Sciences, Beijing 100049, China}

\author[0009-0001-9085-8718]{Jun-Chao Liang}
\affiliation{CAS Key Laboratory of Optical Astronomy, National Astronomical Observatories, Beijing 100101, China}
\affiliation{University of Chinese Academy of Sciences, Beijing 100049, China}

\author[0000-0001-7671-4745]{You-Fen Wang}
\affiliation{CAS Key Laboratory of Optical Astronomy, National Astronomical Observatories, Beijing 100101, China}
\affiliation{University of Chinese Academy of Sciences, Beijing 100049, China}

\author[0000-0001-8869-653X]{Jing Chen}
\affiliation{Nanjing Institute of Astronomical Optics $\&$ Technology, Chinese Academy of Sciences, Nanjing 210042, China}
\affiliation{CAS Key Laboratory of Astronomical Optical $\&$ Technology, Nanjing Institute of Astronomical Optics $\&$ Technology, Nanjing 210042, China}

\author[0000-0003-1454-1636]{Shuo Zhang}
\affiliation{Department of Astronomy, Tsinghua University, Beijing 100084, China}

\author[0000-0002-5818-8769]{Mao-Sheng Xiang}
\affiliation{CAS Key Laboratory of Optical Astronomy, National Astronomical Observatories, Beijing 100101, China}
\affiliation{University of Chinese Academy of Sciences, Beijing 100049, China}

\author[0000-0003-0433-3665]{Hugh R. A. Jones}
\affiliation{School of Physics, Astronomy and Mathematics, University of Hertfordshire, United Kingdom}

\author[0000-0003-3884-5693]{Zhong-Rui Bai}
\affiliation{CAS Key Laboratory of Optical Astronomy, National Astronomical Observatories, Beijing 100101, China}
\affiliation{University of Chinese Academy of Sciences, Beijing 100049, China}

\author[0000-0002-9279-2783]{Xiao-Xiao Ma}
\affiliation{CAS Key Laboratory of Optical Astronomy, National Astronomical Observatories, Beijing 100101, China}
\affiliation{University of Chinese Academy of Sciences, Beijing 100049, China}

\author[0000-0001-7210-0666]{Yun-Jin Zhang}
\affiliation{CAS Key Laboratory of Optical Astronomy, National Astronomical Observatories, Beijing 100101, China}
\affiliation{University of Chinese Academy of Sciences, Beijing 100049, China}

\author[0000-0002-8252-8743]{Hai-Ling Lu}
\affiliation{CAS Key Laboratory of Optical Astronomy, National Astronomical Observatories, Beijing 100101, China}
\affiliation{University of Chinese Academy of Sciences, Beijing 100049, China}

\begin{abstract}

Precise stellar parameters for M-type stars, the Galaxy's most common stellar type, are crucial for numerous studies. In this work, we refined the LAMOST DR10 M-type star catalog through a two-stage process. First, we purified the catalog using techniques including deep learning and color-magnitude diagrams to remove 22,496 non-M spectra, correct 2,078 dwarf/giant classifications, and update 12,900 radial velocities. This resulted in a cleaner catalog containing 870,518 M-type spectra (820,493 dwarfs, 50,025 giants). Second, applying a label transfer strategy using values from APOGEE DR16 for parameter prediction with a ten-fold cross-validated CNN ensemble architecture, we predicted $T_\text{eff}$, $\log g$, [M/H], and [$\alpha$/M] separately for M dwarfs and giants. The average internal errors for M dwarfs/giants are respectively: $T_\text{eff}$ 30/17 K, log $g$ 0.07/0.07 dex, [M/H] 0.07/0.05 dex, and [$\alpha$/M] 0.02/0.02 dex. Comparison with APOGEE demonstrates external precisions of 34/14 K, 0.12/0.07 dex, 0.09/0.04 dex, and 0.03/0.02 dex for M dwarfs/giants, which represents precision improvements of over 20\% for M dwarfs and over 50\% for M giants compared to previous literature results. The catalog is available at \url{https://nadc.china-vo.org/res/r101668/.}

\end{abstract}


\section{Introduction}

M-type stars are the most common stellar type in the Milky Way, comprising over 70\% of the total stellar population \citep{1997AJ....113.2246R,2010AJ....139.2679B,2023A&A...670A..19G}. Based on their mass and evolutionary stage, they can be broadly classified into M dwarfs (dM) and M  giants (gM) \citep{1991ApJS...77..417K}, which exhibit distinct luminosity and radius characteristics in the Hertzsprung-Russell (HR) diagram \citep{2009ssc..book.....G}. The spectra of M-type stars are characterized by strong molecular absorption bands from species such as TiO, VO, and CaH, which complicate the determination of the continuum and make traditional equivalent-width-based measurement methods difficult to apply. These spectral complexities pose significant challenges for the accurate determination of their atmospheric parameters \citep{2015ApJ...804...64M,2019ARA&A..57..571J}.

M dwarf stars are hydrogen-burning stars located at the lower end of the main sequence, typically with a mass less than 0.7 $\text{M}_{\odot}$ and an effective temperature ranging from 2400 K to 4000 K \citep{1993ApJ...402..643K,2023Natur.613..460L}. They have extremely long lifetimes, far exceeding the Hubble time, making them important tools for tracing the chemical and dynamical history of the Milky Way \citep{1997ApJ...482..420L}. In recent years, with the advancement of exoplanet detection technologies, M dwarfs have become preferred targets for the search for Earth-like planets due to their small size and low mass \citep{2016PhR...663....1S,2023A&A...670A..84K,2024MNRAS.531.5053E}. Several potentially habitable exoplanets have been discovered orbiting M dwarfs, including systems such as Proxima Centauri, TRAPPIST-1, and LHS 1140 \citep{2022A&A...658A.115F,2018A&A...613A..68G,2020A&A...642A.121L}. M giants, on the other hand, are high-luminosity, low-temperature stars, including those at the tip of the red giant branch (tRGB), asymptotic giant branch stars (AGB), and red supergiants \citep{1999ApJ...510..232B,2013sse..book.....K}. Due to their high luminosity, M giants are used to trace distant regions and serve as ideal tracers for revealing the accretion and merger events in the Milky Way, especially for studying stellar streams in the Galactic halo, such as the Sagittarius system \citep{1994Natur.370..194I,2016ApJ...823...59L,2019ApJ...886..154Y}. Therefore, to accurately characterize the exoplanetary physical properties of M dwarfs and study the accretion and merger history of the Milky Way, it is essential to precisely measure the fundamental physical parameters of M-type stars, including effective temperature ($T_\text{eff}$), surface gravity (log $g$), and metallicity ([M/H]).

Spectral analysis is a crucial method for obtaining the fundamental parameters of stars, enabling the determination of a star's $T_\text{eff}$, log $g$, [M/H], and other physical characteristics. In recent years, several large-scale survey projects have provided abundant observational data for such studies. For example, high-resolution surveys such as the Apache Point Observatory Galactic Evolution Experiment \citep[APOGEE;][]{2017AJ....154...94M}, the Galactic Archaeology with HERMES \citep[GALAH;][]{2015MNRAS.449.2604D}, the Calar Alto high-Resolution search for M dwarfs with Exo-earths with Near-infrared and optical Echelle Spectrographs \citep[CARMENES][]{2020SPIE11447E..3CQ}, and the Gaia-ESO Survey \citep[GES;][]{2012Msngr.147...25G}; medium-resolution surveys like the Radial Velocity Experiment \citep[RAVE;][]{2006AJ....132.1645S} and the Large Sky Area Multi-Object Fiber Spectroscopic Telescope (LAMOST) Medium-Resolution Spectroscopic Survey \citep{2020arXiv200507210L}; and low-resolution surveys such as the Sloan Extension for Galactic Understanding and Exploration \citep[SEGUE;][]{2009AJ....137.4377Y}, the LAMOST Experiment for Galactic Understanding and Exploration \citep[LEGUE;][]{2012RAA....12..735D}, and the Gaia BP/RP spectra \citep{2023A&A...674A...2D}. These survey projects have made invaluable contributions to stellar spectral research, and the massive datasets from these surveys have created an urgent demand for accurate measurements of M-type star parameters, further driving rapid development in related research.

For the problem of stellar parameter determination from high-resolution spectroscopy of M-type stars, early studies primarily relied on traditional model-fitting techniques. For instance, \cite{2018A&A...620A.180R} were the first to simultaneously use high-resolution optical and near-infrared spectra observed by CARMENES combined with the BT-Settl models \citep{2011ASPC..448...91A}, deriving $T_\text{eff}$, log $g$, and [M/H] for 292 M dwarfs through least-squares fitting, with uncertainties of 100 K, 0.3 dex, and 0.3 dex, respectively. \cite{2018A&A...615A...6P,2019A&A...627A.161P} employed the updated PHOENIX-ACES models \citep{2013A&A...553A...6H} and $\chi^2$ minimization methods to determine $T_\text{eff}$, log $g$, and [M/H] for 300 M dwarfs, improving the uncertainties to 51 K, 0.07 dex, and 0.16 dex. \cite{2021A&A...656A.162M} derived atmospheric parameters for 343 M dwarfs based on CARMENES spectra employing the STEPARSYN Bayesian spectral synthesis implementation, combining BT-Settl models with the radiative transfer code \texttt{turbospectrum}, and imposed Bayesian priors based on multi-band photometric data to avoid any potential degeneracy in the parameter space. In recent years, data-driven methods have rapidly developed. \cite{2020A&A...642A..22P} first applied convolutional neural networks (CNN) trained on synthetic PHOENIX-ACES spectra to directly predict $T_\text{eff}$, log $g$, [M/H], and $v\sin{i}$ from dM spectra observed by CARMENES, but encountered a significant ``synthetic gap'' problem, where differences in feature distributions between synthetic and observed spectra led to notable biases in metallicity predictions. To mitigate this issue, \cite{2023A&A...673A.105B} developed a deep transfer learning approach, utilizing $T_\text{eff}$ determined from interferometry and [M/H] from FGK + M binaries to fine-tune the internal features of neural networks trained on synthetic PHOENIX-ACES spectra, achieving improvements of 20 K and 0.2 dex in $T_\text{eff}$ and [M/H] accuracy, respectively, compared to \cite{2020A&A...642A..22P}. \cite{2024A&A...687A.205M} further proposed a feature-based deep transfer learning method that does not require high-quality measurements of parameters involved in the knowledge transfer, effectively alleviating the impact of the ``synthetic gap''.

However, extending the successful experience of M-type stellar parameter measurements from high-resolution spectroscopy to low-resolution large-scale surveys faces even more formidable challenges. Compared to G-type stars, the dominance of molecular lines in M-type stellar atmospheres results in more pronounced discrepancies between theoretical models and actual observations. In the optical band, only about half of the spectral lines have accurately determined laboratory parameters \citep{2014dapb.book...39K}, and the abundance of molecular features in cool stellar atmospheres further increases the complexity of modeling \citep{2019ARA&A..57..571J}. These theory-observation inconsistencies become even more pronounced in low-resolution spectra, directly compromising the accuracy of parameter measurements. For instance, the LAMOST Stellar Parameter Pipeline for M-type stars (LASPM) developed by \cite{2021RAA....21..202D}, which finds the best-matching templates in the BT-Settl synthetic spectral library for each observed spectrum and then minimizes $\chi^2$ through a linear combination of the five best-matching templates to derive atmospheric parameters, exhibits significant systematic deviations when compared to high-resolution APOGEE results, such as an overestimation of surface gravity by approximately 0.63 dex and an underestimation of metallicity by about 0.25 dex, along with considerable dispersion exceeding 0.2 dex for both surface gravity and metallicity. These discrepancies highlight the limitations of applying theoretical models to the measurement of M-type stellar parameters from low-resolution spectroscopy.

To address the challenges of low-resolution M-type spectral analysis, researchers have explored various approaches. \cite{2021ApJS..253...45L} employed the SLAM model to transfer the high-precision labels from APOGEE DR16 to LAMOST spectra, measuring $T_\text{eff}$ and [M/H] for approximately 300,000 dMs. \cite{2023RAA....23e5008Q} applied the SLAM method to gMs, deriving $T_\text{eff}$, log $g$, [M/H], and alpha-element abundance ([$\alpha$/M]) for 43,972 gM spectra. \cite{2022ApJS..260...45D} used the MILES \citep{2011A&A...532A..95F} empirical spectral library to determine $T_\text{eff}$, log $g$, and [M/H] for 763,136 spectra of M-type stars in LAMOST DR8. \cite{2024ApJS..275....8L} proposed a spectral emulator approach using the MaStar observed spectral library to measure $T_\text{eff}$, log $g$, [M/H], and [$\alpha$/M] for O- to M-type stars in LAMOST DR10. \cite{2024ApJS..275...42D} improved the accuracy of LASPM parameter measurements by constructing a more self-consistent M-type star spectral library. \cite{2025ApJS..277...47Z} applied the Cycle-StarNet technique to significantly reduce the discrepancies between theoretical and observed spectra and derived $T_\text{eff}$, log $g$, and [M/H] for 507,513 dM spectra in LAMOST DR10. Most of these studies adopted parameter inference methods based on iterative optimization, compared to APOGEE results, the dispersion in log $g$ and [M/H] for M giants is generally larger than that for M dwarfs, typically in the range of 0.2 dex to 0.32 dex. Therefore, further improving the measurement precision of log $g$ and [M/H] for M-type stars, particularly for M giants, remains a key focus for future research.

This work adopted a ``label transfer + parameter prediction'' strategy to estimate M-type star parameters, where a neural network model is trained using LAMOST spectra with high-precision parameters from APOGEE and then applied to other LAMOST spectra. This approach effectively reduces systematic errors introduced by the ``synthetic gap'' \citep{2020A&A...642A..22P}, relative to directly applying synthetic-trained models to observed spectra. Compared to the iterative optimization-based parameter inference methods commonly used in previous studies, the ``parameter prediction'' method employed in this work improves both efficiency and accuracy in parameter estimation. The structure of this paper is as follows. Section~\ref{sec:Data} describes the data used in this work. Section~\ref{sec:Method} introduces the workflow, spectral preprocessing, CNN framework, and technical details. Section~\ref{sec:Result} presents the construction of the Recommended Catalog and a comparative analysis of the stellar parameters provided in this work with those from previous studies. The conclusions are summarized in Section~\ref{sec:Summary}.

\section{Data}
\label{sec:Data}
\subsection{LAMOST}
LAMOST (also called the Guo Shou Jing Telescope), located in the Xinglong station of National Astronomical Observatories, Chinese Academy of Sciences (NAOC), is a special reflecting Schmidt telescope with 4000 fibers in a field of view of 20 $\rm deg^{2}$ in the sky \citep{1996ApOpt..35.5155W, 2004ChJAA...4....1S, 2012RAA....12.1197C}. LAMOST is dedicated to a spectral survey of celestial objects over the entire available northern sky, providing a large number of medium ($R$ $\sim$ 7500) and low resolution ($R$ $\sim$ 1800) spectra \citep{2012RAA....12..735D, 2012RAA....12..723Z}. The regular survey of LAMOST started in September 2012 \citep{2015RAA....15.1095L}, and as of DR10 v1.0\footnote{Data available at \url{https://www.lamost.org/dr10/v1.0/}. We chose to use DR10 v1.0 rather than the later-released DR10 v2.0 because: (1) our analysis was initiated using v1.0, and was already in its final stages when v2.0 became available, and (2) v1.0 contains a larger dataset that enables us to construct a larger M-type star catalog.}, LAMOST published over 11.8 million low resolution spectra with a wavelength range of 3700--9000 \AA, which includes about 11.4 million stellar spectra. Additionally, LAMOST DR10 v1.0 also provides a Catalog of M-type stars, which contains 818,676 dM spectra, 53,481 gM spectra, and 3,977 M subdwarf (sdM) spectra, and all of this work was conducted based on this catalog.

\subsection{Gaia}
The European Space Agency’s (ESA) Gaia mission \citep{2016A&A...595A...1G}, initiated in 2013, has progressed to its third data release (Gaia DR3). The Gaia satellite was equipped with three main instruments: the astrometric instrument to collect images in the broad $G$-band (330--1050 nm), the blue BP and red RP prism photometers for low-resolution spectra, and the Radial Velocity Spectrometer (RVS) for measuring radial velocities (RV) \citep{2023A&A...674A...1G}. Gaia DR3 provides $G$-band photometry for 1.8 billion sources, parallax, proper motion, $G_{\rm BP}$ and $G_{\rm RP}$ for 1.5 billion sources, and RV measurements for 33 million bright stars ($G_{\rm RVS}<14$, $3100<T_{\rm eff}<14500$ K) \citep{2023A&A...674A..28F}. Additionally, it also provides the Re-normalised Unit Weight Error \citep[RUWE;][]{2018A&A...616A...2L}, a metric designed to assess the goodness of the astrometric fit.

\subsection{APOGEE}
The Sloan Digital Sky Survey \citep[SDSS;][]{2000AJ....120.1579Y} started observations in 1998 and now is in its fourth phase\citep[SDSS-IV;][]{2017AJ....154...28B}. The collected data of SDSS-IV includes optical images of most of the northern high Galactic latitude sky as well as optical and near infrared spectroscopy of over 3.5 million targets, and these observations all used the 2.5 m Sloan Foundation Telescope at Apache Point Observatory \citep{2006AJ....131.2332G}. APOGEE-2 is a key project of SDSS-IV, and a major near-infrared (15140--16940 \AA) spectroscopic survey to investigate the composition and dynamics of stars in the Galaxy, based on multiplexed high-resolution ($R$ $\sim$ 22,500) spectrographs. APOGEE DR16\footnote{To ensure complete parameter coverage, we used APOGEE DR16 rather than DR17, as the latter excludes [M/H] and [$\alpha$/M] values for M dwarf stars with $T_\text{eff}$ below 3500 K due to known systematic issues \citep{2022ApJS..259...35A}.} observed spectra for about 430,000 stars of both the northern and southern sky\citep{2020AJ....160..120J}, and provided atmospheric parameters ($T_{\rm eff}$, log $g$, [M/H]) and chemical abundances (such as [$\alpha$/M], [C/M], and [N/M]) using the APOGEE Stellar Parameter and Chemical Abundance Pipeline \citep[ASPCAP;][]{2016AJ....151..144G}, which was developed for the automated analysis of observed spectra.

\subsection{Spectral Templates} \label{sec:spectra_templates}
In this work, when employing the template matching method to measure the RVs of M-type stars, spectral templates from \cite{2016IAUS..317..371Z} and LAMOST spectral analysis pipeline \citep[LAMOST 1D Pipeline;][]{2012RAA....12.1243L} were used. \cite{2015AJ....150...42Z} and \cite{2015RAA....15.1154Z} detailed the development of templates for M dwarf and giant stars, respectively. The dwarf star templates are derived from the SDSS DR7 M-type star catalog published by \cite{2011AJ....141...97W}, while the giant star templates were selected from M-type star spectra of LAMOST DR1. \citeauthor{2016IAUS..317..371Z}’s catalog contains 223 M-type star templates, including only seven giant star templates covering a wavelength range of 4700--9000 \AA. The LAMOST 1D Pipeline provides 10 dwarf (dM0--dM9) and 10 giant (gM0--gM9) templates, with a wavelength range of 3865--8997 \AA. Due to the limited number of giant templates in \citeauthor{2016IAUS..317..371Z}'s collection, which are significantly fewer than the dwarf templates and lack late-type giants beyond M6, this work combined \citeauthor{2016IAUS..317..371Z}’s templates with all 10 giant templates from the LAMOST 1D pipeline for measuring RVs. The resolution of the M-type star templates from both \cite{2016IAUS..317..371Z} and LAMOST is $\sim$ 1800. Spectral preprocessing, such as wavelength range alignment, was conducted during the merging of the templates, as detailed in Section~\ref{sec:preprocess}.

\section{Method}
\label{sec:Method}
\subsection{Workflow}
\label{sec:workflow}

\begin{figure*}
	\centering
	\includegraphics[width=1\linewidth]{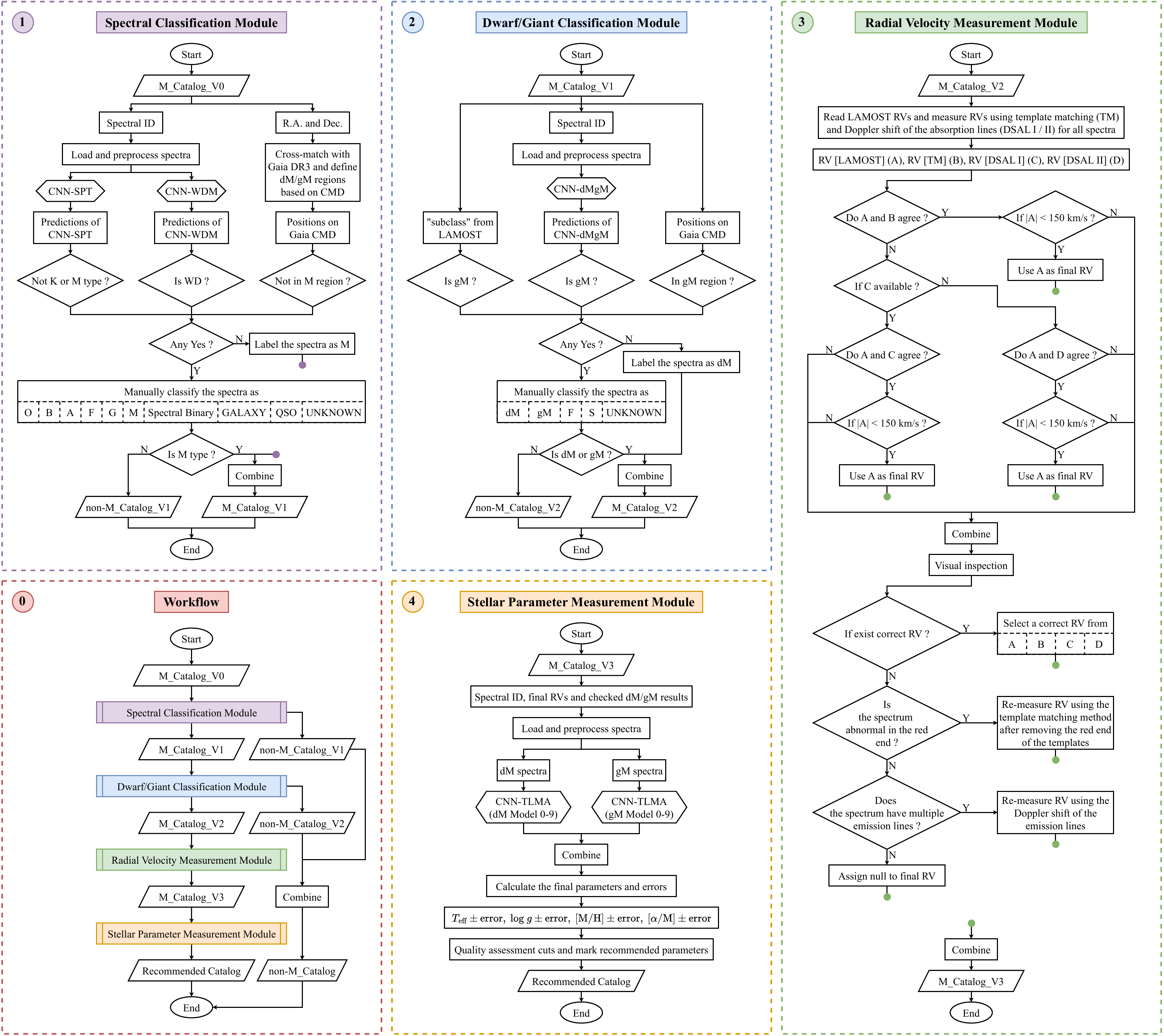}
	\caption{Workflow diagram. Panel~0 illustrates the workflow for constructing a purer M-type star catalog, which consists of four modules: the Spectral Classification Module (SCM), the Dwarf/Giant Classification Module (DGCM), the Radial Velocity Measurement Module (RVMM), and the Stellar Parameter Measurement Module (SPMM). These modules are designed for spectral classification, M dwarf (dM) and M giant (gM) classification, radial velocity (RV) measurement, and stellar parameter estimation, respectively. Panels~1 to 4 present the flowcharts of these four modules. It is important to note that in Panels~1 and 3, purple and green dots are used to indicate breakpoints to avoid overlap between connecting lines. In reality, these breakpoints lie along the same connecting line.
	}
	\label{workflow_diagram}
\end{figure*}

In the M-type star catalog of LAMOST DR10 (we called M\_Catalog\_V0 hereafter), there are a total of 876,134 spectra, including 822,653 dwarf star spectra and 53,481 giant star spectra. However, the classification results from the LAMOST 1D Pipeline are not entirely accurate, with some non-M-type stars mixed in, and errors in the classification between dwarf and giant stars for some spectra. Additionally, the RVs provided by LAMOST also contain errors in some cases. To obtain a more reliable M-type star catalog (we called Recommended Catalog hereafter), with more accurate spectral type, dwarf/giant classification, RV, and stellar parameter measurements, we designed a specific procedure to achieve as much purity as possible. 

To construct the Recommended Catalog, we designed four modules: the Spectral Classification Module (SCM), the Dwarf/Giant Classification Module (DGCM), the Radial Velocity Measurement Module (RVMM), and the Stellar Parameter Measurement Module (SPMM). In SCM, DGCM and SPMM modules, we trained multiple one-dimensional convolutional neural network \citep[1D CNN;][]{2020PASP..132b4504Z} models using PyTorch \citep{Paszke2019PyTorchAI} to classify spectra and measure stellar parameters of LAMOST M-type stars. In the SCM, we trained a model for spectral classification of OB (since the number of O- and B-type stars is relatively small, we combine them into a single class, abbreviated as OB), A, F, G, K, M, GALAXY, and QSO (CNN-SPT), as well as a model for white dwarf (WD) and M spectra classification (CNN-WDM). In the DGCM, we developed a model for classifying dM and gM spectra (CNN-dMgM) to correct potential luminosity classification errors in LAMOST M-type stars. In the RVMM, we adopted two methods to measure the RVs of M-type stars: template matching and Doppler shift methods. In the SPMM, we trained models to measure parameters of $T_\text{eff}$, log $g$, [M/H], [$\alpha$/M] (CNN-TLMA), aiming to address issues in parameter measurements of LAMOST M-type stars and provide additional stellar parameters.

Figure~\ref{workflow_diagram} shows the workflow (Panel~0) of this work and the detailed procedures (Panels~1--4) of the four modules, where each module progressively enriched the output catalog by incorporating newly derived useful information generated through the processing. First, the SCM refined the spectral classification by identifying and removing non-M-type stars from M\_Catalog\_V0. The confirmed M-type stars were stored in M\_Catalog\_V1, while the rejected non-M-type stars were collected in non-M\_Catalog\_V1. Second, the DGCM processed M\_Catalog\_V1 to improve the dwarf/giant classification accuracy. During this step, additional non-M-type stars that were missed by SCM were identified and stored in non-M\_Catalog\_V2, while the remaining M-type stars with updated dM/gM classifications were saved as M\_Catalog\_V2. Third, the RVMM processed M\_Catalog\_V2 to provide corrected RV measurements, updating the catalog to M\_Catalog\_V3. Finally, the SPMM predicted stellar parameters and corresponding errors based on the updated RV results in M\_Catalog\_V3, generating the Recommended Catalog. Since no additional non-M-type stars were removed after the DGCM stage, the non-M-type stars from non-M\_Catalog\_V1 and non-M\_Catalog\_V2 were combined to form the comprehensive non-M\_Catalog as a byproduct of this purification process. After passing through these four modules, 5,616 non-M-type spectra were removed, the dM/gM classification for 2,078 spectra was corrected, RVs for 12,900 spectra were revised, and the parameters $T_\text{eff}$, log $g$, [M/H], and [$\alpha$/M] were measured for 870,518 spectra.

It should be noted that we previously identified 16,880 non-M-type spectra in the DR8 M-type star catalog and reclassified them. According to the data release requirements at that time, subsequent data releases after DR8 used our classification results to update these spectra to the correct types, and spectra classified as UNKNOWN due to low signal-to-noise ratio or data quality issues were no longer included in releases after DR8. Therefore, the final non-M\_Catalog does not include the 16,880 spectra whose classifications had already been updated in the LAMOST official releases, and it includes only the 5,616 non-M-type spectra newly identified from the ninth and tenth observing years. Our refined M-type star catalog---the Recommended Catalog---also excludes all of these identified non-M-type spectra.

\subsection{Spectral Preprocessing}
\label{sec:preprocess}

As described in Section~\ref{sec:workflow}, we designed four modules in this work: SCM, DGCM, RVMM, SPMM, and performed preprocessing on the input spectra. Below are the preprocessing steps involved in this work, which vary across each module and will be detailed further in Section~\ref{sec:Mudules}. 
\begin{enumerate}

    \item Remove the flux points corresponding to the \texttt{ormask}\footnote{The \texttt{ormask} is a decimal integer represented by a six-bit binary number for each wavelength of the LAMOST spectrum. A value of 0 indicates that no issues were encountered in any exposure during the spectrum reduction process, while a non-zero value means that at least one exposure encountered issues during the spectrum reduction process for that wavelength.\label{foot:ormask}} value of non-zero. \label{item:ormask}

    \item Shift the spectra to the rest frame. \label{item:rmrv}

    \item Linearly interpolate the spectra to have a wavelength interval of 1 \AA. \label{item:interp}

    \item Reset the wavelength range of the spectra to
    \begin{enumerate}[label*=\alph*]
        \item 3900--8899 \AA \label{item:wave3900}
        \item 4700--8899 \AA \label{item:wave4700}
    \end{enumerate}

    \item Z-score Normalization \label{item:norm}
    \begin{equation*}
    \frac{x - x_{\mu}}{x_{\sigma}}
    \end{equation*}

\end{enumerate}

In the SCM and DGCM modules, the preprocessing steps used for the CNN-SPT, CNN-WDM, and CNN-dMgM models include operations \ref{item:ormask}, \ref{item:interp}, \ref{item:wave3900}, and \ref{item:norm}. In the RVMM module, when measuring RVs using the template matching method, only step \ref{item:ormask} was used for preprocessing the observed spectra, while steps of \ref{item:interp}, \ref{item:wave4700}, and \ref{item:norm} were used to preprocess the template spectra. This ensures that both \citeauthor{2016IAUS..317..371Z}'s templates and the 10 templates of LAMOST 1D pipeline (mentioned in Section~\ref{sec:spectra_templates}) have consistent wavelength ranges and flux scales. For measuring RVs using the Doppler shift method, no preprocessing was performed on the observed spectra. In the SPMM module, the preprocessing steps used for the CNN-TLMA model include \ref{item:ormask}, \ref{item:rmrv}, \ref{item:interp}, \ref{item:wave3900}, and \ref{item:norm}, where the step \ref{item:rmrv} used RVs from the RVMM module.

\subsection{CNN Framework}

\begin{figure}
	\centering
	\includegraphics[width=0.8\linewidth]{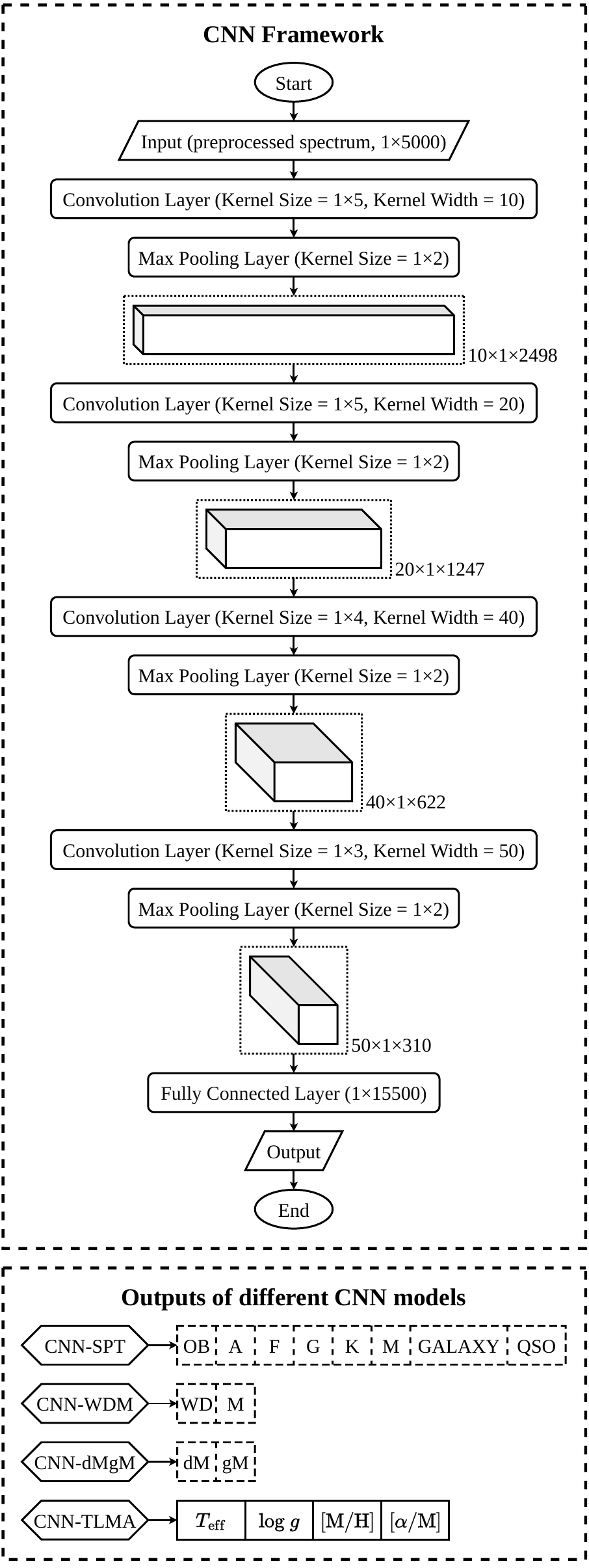}
	\caption{The CNN network architecture used in this work and the outputs of the different CNN models. The CNN-SPT model is used for spectral type classification, the CNN-WDM model is used for white dwarf (WD) and M-type star classification, the CNN-dMgM model is used for dM and gM classification, and the CNN-TLMA model is used for measuring the stellar parameters. It should be noted that the outputs of CNN-SPT, CNN-WDM, and CNN-dMgM are the probabilities of each type, while the output of CNN-TLMA consists of four stellar parameters.}
	\label{CNN_framework}
\end{figure}

As described in Section~\ref{sec:workflow}, we designed four CNN models in this work: CNN-SPT, CNN-WDM, CNN-dMgM, and CNN-TLMA, where the first three are classification models, and the last one is a regression model. The common feature of all four models is that they share the same network architecture, as shown in Figure~\ref{CNN_framework}, which consists of four convolutional layers, four max-pooling layers, and one fully connected layer. The activation function used in each layer was Rectified Linear Unit (ReLU), the optimizer was Adam \citep{Kingma2014AdamAM}, the batch size was set to 128, and the initial learning rate was set to 0.001. An adaptive learning rate adjustment method was used to help the model escape local minima, and training was terminated early to prevent overfitting when the validation loss reaches its minimum. The key differences between the models lie in the labels and the loss functions. For the classification models (CNN-SPT, CNN-WDM, CNN-dMgM), the labels were one-hot encoded, and the loss function used was cross-entropy. The softmax function was applied to convert the model's output into class probabilities, and the class with the highest probability was taken as the predicted spectral type. In contrast, for the regression model (CNN-TLMA), the labels were Z-score normalized to eliminate scale differences between the labels. The loss function used was mean squared error, and the model's output was denormalized to return the predicted stellar parameters on their original scale.

\subsection{Module Analysis}
\label{sec:Mudules}
\subsubsection{Spectral Classification Module}
\label{sec:SCM}

\begin{figure}
	\centering
	\includegraphics[width=0.9\linewidth]{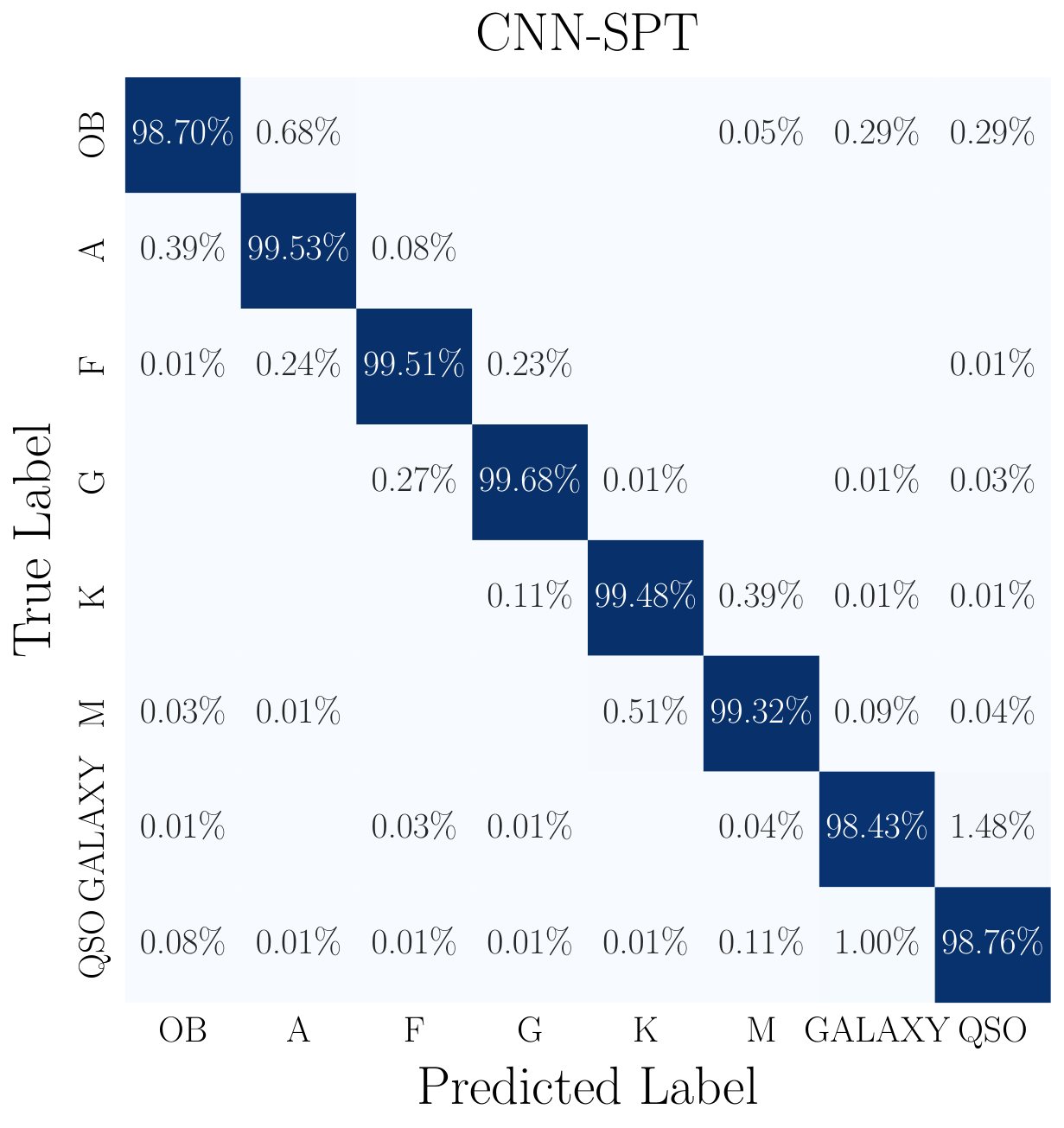}
	\caption{Confusion matrix of the CNN-SPT model on the validation set.}
	\label{fig:CM_SPT}
\end{figure}

\begin{figure}
	\centering
	\includegraphics[width=0.5\linewidth]{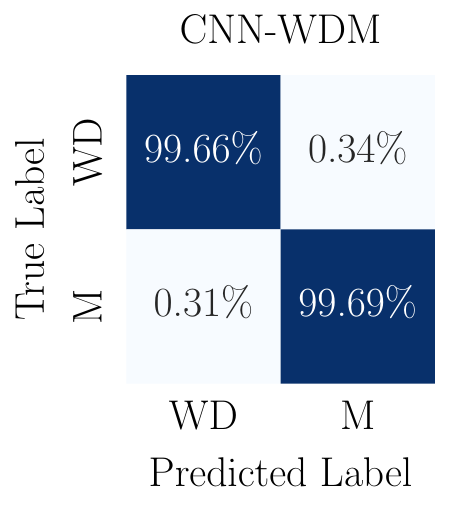}
	\caption{Confusion matrix of the CNN-WDM model on the validation set.}
	\label{fig:CM_WDM}
\end{figure}

\begin{figure}
	\centering
	\includegraphics[width=1\linewidth]{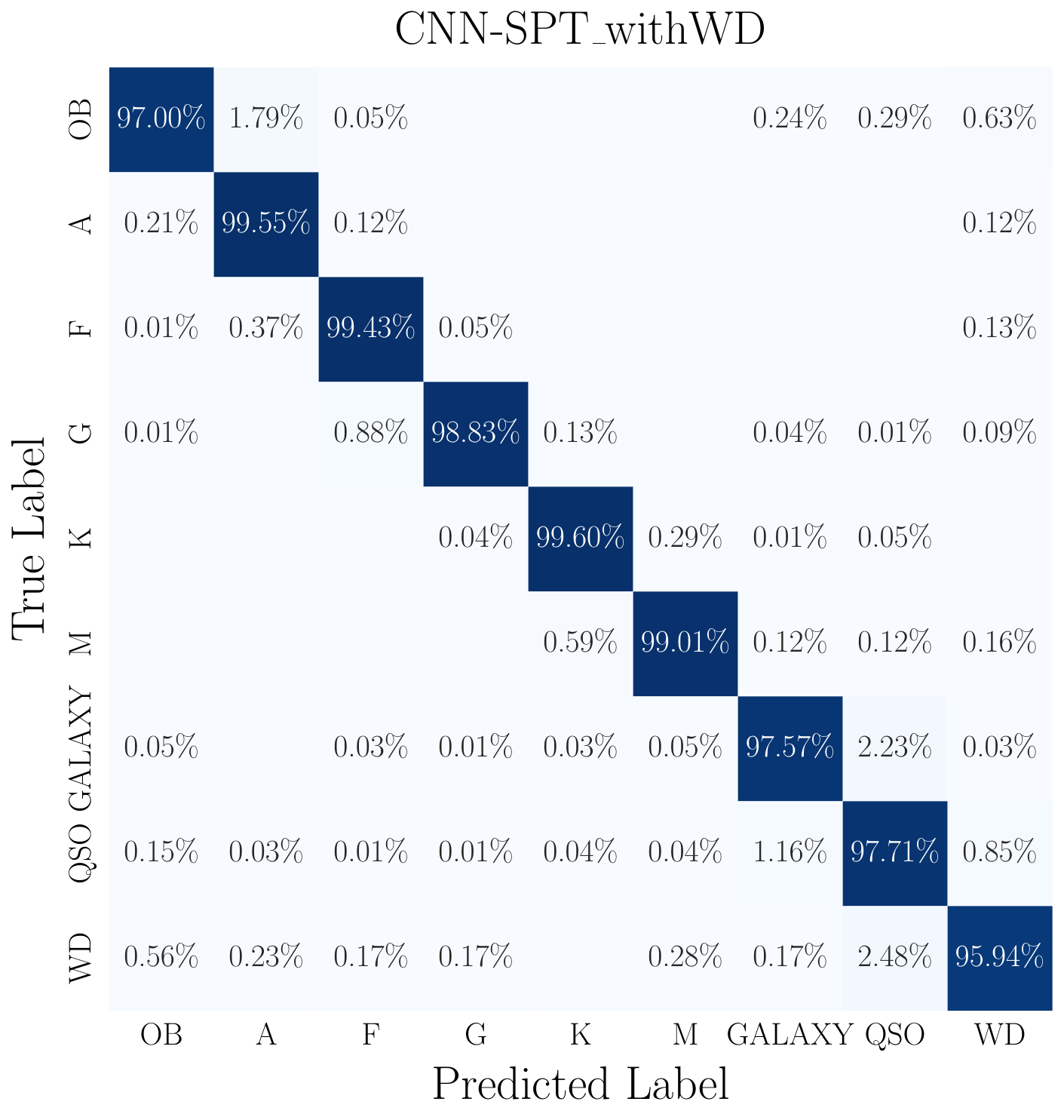}
	\caption{Confusion matrix of the CNN-SPT model with WD type included on the validation set.}
	\label{fig:CM_SPT_withWD}
\end{figure}

To identify non-M-type spectra from M\_Catalog\_V0 as thoroughly as possible, we combined the prediction results from CNN-SPT and CNN-WDM models, as well as the positions on the Gaia Color-Magnitude Diagram (CMD), to select non-M-type candidates discussed later. First, we trained two CNN models---CNN-SPT and CNN-WDM---to classify the spectra in the M\_Catalog\_V0. Both the training data and labels are from the LAMOST DR10 LRS Catalogs\footnote{\url{https://www.lamost.org/dr10/v1.0/catalogue}}. For simplicity, we will refer to these catalogs by their names without the prefix ``LAMOST LRS'' in subsequent descriptions. The spectral types include OB, GALAXY, QSO from the ``General Catalog'', A, F, G, K from the ``Stellar Parameter Catalog of A, F, G, and K Stars'', M from the ``Catalog of gM, dM, and sdM Stars'' (i.e., M\_Catalog\_V0), and WD from the ``Catalog of White Dwarf Stars''. To ensure the reliability of the training data and labels, we excluded spectra with S/N\_max $<$ 5, where S/N\_max = max(S/N\_$u$, S/N\_$g$, S/N\_$r$, S/N\_$i$, S/N\_$z$) represents the maximum signal-to-noise ratio (S/N) across the $u$, $g$, $r$, $i$, and $z$ bands. This criterion ensures that each retained spectrum has at least one band with sufficient spectral quality for analysis. Such low-S/N\_max spectra account for only 2.90\% of all LAMOST-released spectra. Additionally, we cross-referenced the ``golden sample of OBA and FGKM stars'' from \cite{2023A&A...674A..39G}, the ``good QSO and galaxy candidates'' from \cite{2023A&A...674A...1G}, and the SIMBAD astronomical database \citep{2000A&AS..143....9W} using TOPCAT \citep{2005ASPC..347...29T} software and a $3^{\prime\prime}$ matching radius. All subsequent catalog cross-matching was performed using the same tool and matching radius. We filtered the labels based on both the sample size and the consistency of classification results. From the filtered samples, we randomly selected 30,000 spectra from each category as the final training data. The training and validation sets were split in a 3:1 ratio. Since the number of OB and WD spectra was insufficient (8279 and 7094, respectively), we used weighted loss calculation to address the data imbalance issue. The weight values assigned to OB and WD were 3.62 and 4.23, respectively, while all other categories were assigned a weight of 1. Figures \ref{fig:CM_SPT} and \ref{fig:CM_WDM} show the confusion matrices for CNN-SPT and CNN-WDM on the validation set, with average accuracies of 99.18\% and 99.68\%, respectively. If WD is included as an output type in the CNN-SPT model, the confusion matrix on the validation set, as shown in Figure~\ref{fig:CM_SPT_withWD}, results in an average accuracy of only 98.29\%, with the accuracy for WD dropping from 99.69\% to 95.94\%. Therefore, we adopted a two-model approach for spectral type classification. The CNN-SPT model classifies the spectra from M\_Catalog\_V0 into nine categories: OB, A, F, G, K, M, GALAXY, QSO, with 3,327 spectra predicted as non-K or non-M types. The CNN-WDM model classifies the spectra into two categories: WD and M, with 3,809 spectra predicted as WD. It is important to emphasize that although the spectral types of inputs to CNN-SPT and CNN-WDM may potentially fall outside the predefined output categories of these models, we conduct rigorous manual verification of all identified non-M-type candidate spectra to ensure the reliability of our classification methodology.

\begin{figure}
	\centering
	\includegraphics[width=1\linewidth]{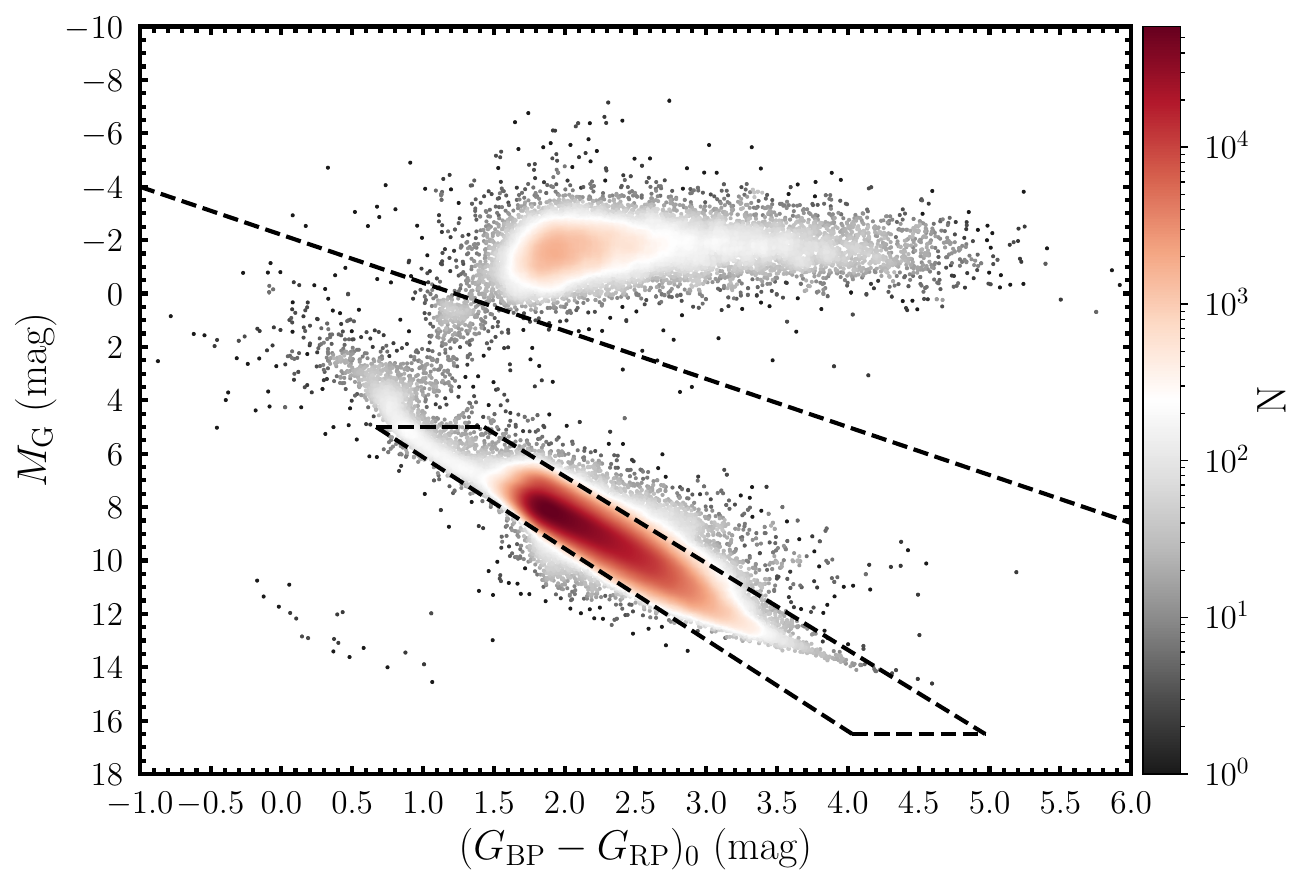}
	\caption{The color-magnitude diagram of the M-type star catalog (M\_Catalog\_V0) of LAMOST DR10. The Gaia absolute magnitude and color used in this diagram are extinction-corrected values. The quadrilateral enclosed by the dashed line marks the dM region, with the coordinates of the four vertices being (0.7, 5), (1.4, 5), (4, 16.5), and (5, 16.5). The gM region is separated by the dashed line above, which satisfies the equation $M_\mathrm{G}=1.8\times(G_\mathrm{BP}-G_\mathrm{RP})_{0}-2.2$.}
    \label{CMD_all_criterion}
\end{figure}

Next, we filtered samples not in the dM or gM regions on the Gaia CMD. To obtain Gaia's color and magnitude data, we cross-matched the M\_Catalog\_V0 with Gaia DR3, yielding 658,857 sources and 866,842 spectra. Figure~\ref{CMD_all_criterion} shows the extinction-corrected Gaia color-magnitude diagram, where the absolute magnitude $M_\text{G}$ for each star is calculated using the following formula:

\begin{equation}
\label{eq:M_G}
M_\text{G} = G - 5\lg(d) + 5 - \text{A}_G
\end{equation}

where the distance $d$ is taken from \cite{2021AJ....161..147B}'s \texttt{r\_med\_geo}, the reddening $ \text{E}(B-V) $ is determined from the 3D dust map Bayestar19 \citep{2019ApJ...887...93G}, and the extinction coefficient is from \cite{2018MNRAS.479L.102C}. We used the criteria mentioned in Section~3.1 of \cite{2021ApJS..253...45L} to select spectra with reliable Gaia colors $G_\text{BP} - G_\text{RP}$ and apparent magnitudes $G$, and the selection criteria are as follows.

\begin{enumerate}
    \item $\texttt{parallax\_over\_error} > 5$
    \item $\texttt{phot\_g\_mean\_flux\_over\_error} > 20$
    \item $\texttt{phot\_bp\_mean\_flux\_over\_error} > 20$
    \item $\texttt{phot\_rp\_mean\_flux\_over\_error} > 20$
    \item $\texttt{ruwe} < 1.4$
\end{enumerate}

\begin{table*}
	\normalsize
	\centering
	\caption{The visual inspection results of 13,436 non-M-type star candidate spectra.}
	\begin{tabular*}{\linewidth}{@{\extracolsep{\fill}}cccccccccc}
		\toprule
		\toprule
		Spectral type & B & A & F & G & M & Spectral Binary & GALAXY & QSO & UNKNOWN\\
		\midrule
        Number & 1 & 3 & 1 & 5 & 9530 & 484 & 73 & 18 & 3321\\
		\bottomrule
	\end{tabular*}
	\label{tab:sptype_result}
\end{table*}

\begin{table*}
	\normalsize
	\centering
	\caption{The visual inspection results of 16,880 non-M-type spectra in the LAMOST DR8 M-type star catalog.}
	\begin{tabular*}{\linewidth}{@{\extracolsep{\fill}}ccccccccccc}
		\toprule
		\toprule
		Spectral type & O & B & A & F & G  & Spectral Binary & EM Star & GALAXY & QSO & UNKNOWN\\
		\midrule
        Number & 1 & 115 & 14 & 49 & 16 & 45 & 54 & 45 & 46 & 16495\\
		\bottomrule
	\end{tabular*}
	\label{tab:sptype_noM_dr8}
\end{table*}

\begin{figure*}
	\centering
	\includegraphics[width=1\linewidth]{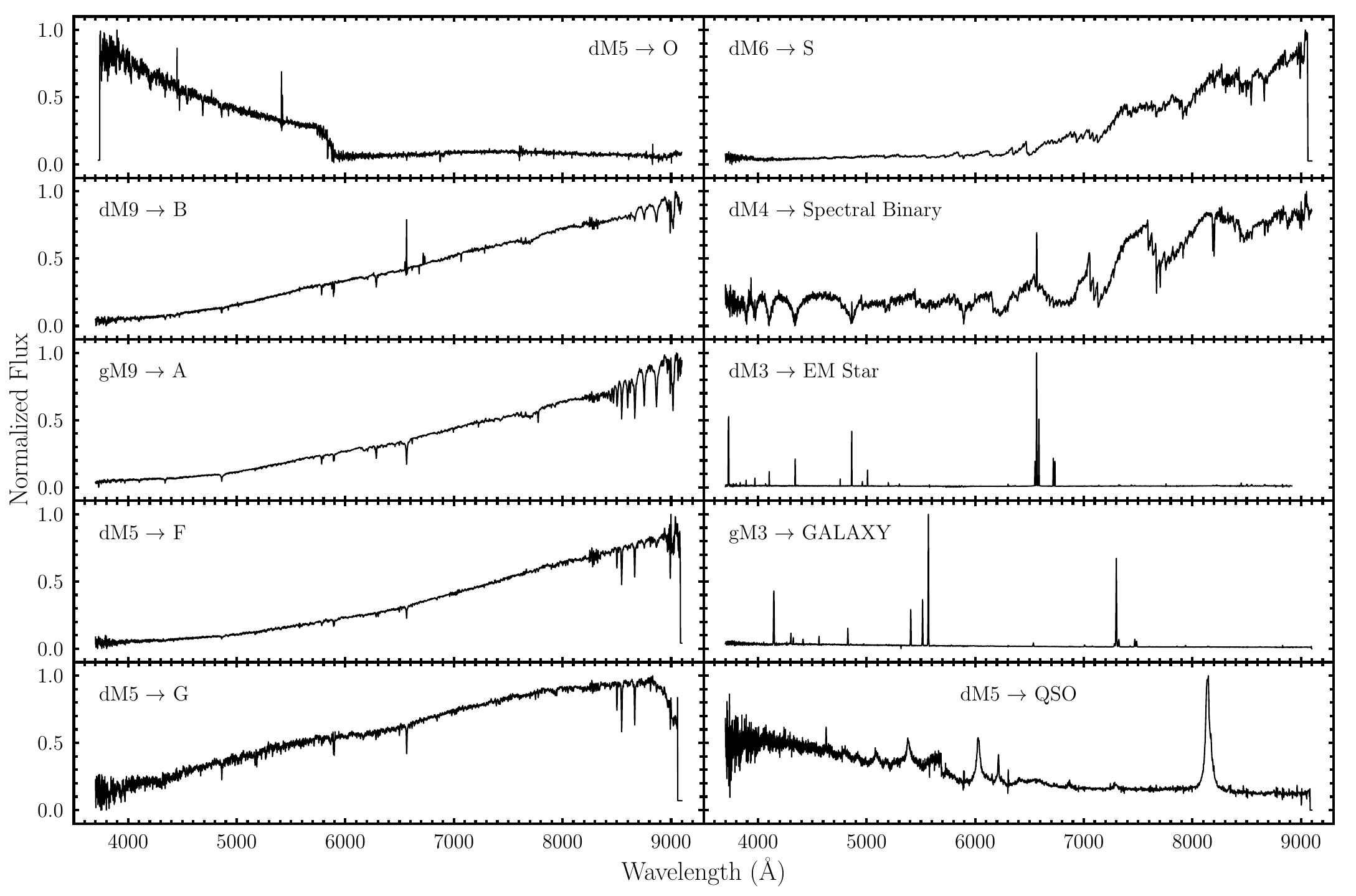}
	\caption{The examples of non-M-type spectra identified in this work. In all subfigures, the labels on the left and right of the arrows represent the spectral types published by LAMOST and the spectral types corrected in this work, respectively.}
	\label{demo_spectrum_noM}
\end{figure*}

Based on Figure~1 in \cite{2021ApJS..253...45L}, we outlined the dM region in quadrilateral shapes in Figure~\ref{CMD_all_criterion}, and a line was empirically used to separate the gM region. A total of 7,886 spectra in Figure~\ref{CMD_all_criterion} fall outside the dM and gM regions. We combined three sets of spectra: those predicted by CNN-SPT as non-K or non-M, those predicted by CNN-WDM as non-M, and those not in the dM or gM regions on the Gaia CMD, resulting in 13,436 spectra for visual inspection. The results of the visual inspection are shown in Table~\ref{tab:sptype_result}, with 3,906 non-M-type spectra identified and stored in the non-M\_Catalog\_V1. The remaining 9,530 spectra classified as M-type and the 862,698 spectra directly marked as M were combined into a total of 872,228 spectra, which were stored in the M\_Catalog\_V1. As mentioned in Section~\ref{sec:workflow}, we also corrected the types of 16,880 non-M-type spectra found in DR8, and the updated spectral types are shown in Table~\ref{tab:sptype_noM_dr8}. Figure~\ref{demo_spectrum_noM} presents examples of the non-M-type spectra we identified, it also shows the changes in their classification results. Here, EM Stars are non-M type stars with emission lines in their spectra.

\subsubsection{Dwarf/Giant Classification Module}
\label{sec:DGCM}

\begin{figure}
	\centering
	\includegraphics[width=0.5\linewidth]{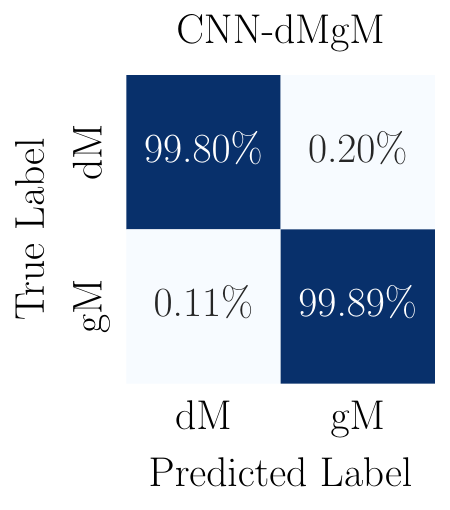}
	\caption{Confusion matrix of the CNN-dMgM model on the validation set.}
	\label{fig:CM_dMgM}
\end{figure}

\begin{table*}
	\normalsize
	\centering
    \begin{threeparttable}
        \caption{Criteria\tnote{a} for manually classifying M dwarf (dM) and M giant (gM).}
    	\begin{tabular*}{\textwidth}{@{\extracolsep{\fill}}ccccc}
    		\toprule
    		\toprule
    		 Feature & CaH  & K \scriptsize I \normalsize doublet & Na \scriptsize I \normalsize doublet & Ca \scriptsize II \normalsize triplet\\
    		\midrule
    		$\lambda$\tnote{b} (\AA) & 6946$-$7050  & 7665, 7699 & 8183, 8195 & 8498, 8542, 8662\\
    		\midrule
            dM & dip & deep & deep & shallow \\
            gM & flat & shallow & shallow & deep \\
    		\bottomrule
    	\end{tabular*}
        \label{tab:dMgM_check_feature}
    	\begin{tablenotes}
    		\item[a] The classification criteria for dM and gM was summarized from \cite{1991ApJS...77..417K}, \cite{2012ApJ...753...90M}, and \cite{2015RAA....15.1182G}.
    		\item[b] The air wavelengths of the features are from Table~5 in \cite{1991ApJS...77..417K}. Since the LAMOST spectra are in vacuum wavelengths, for consistency, we converted the air wavelengths of the features to vacuum wavelengths during the manual inspection of the spectra.
    	\end{tablenotes}
    
    \end{threeparttable}
\end{table*}

\begin{table*}
	\normalsize
	\centering
	\caption{The results of the manual inspection of 53,987 gM candidate spectra.}
	\begin{tabular*}{\textwidth}{@{\extracolsep{\fill}}cccccc}
		\toprule
		\toprule
		Spectal type & dM & gM & F & S & UNKNOWN\\
		\midrule
        Number & 2671 & 50025 & 2 & 85 & 1204\\
		\bottomrule
	\end{tabular*}
	\label{tab:dMgM_result}
\end{table*}

\begin{figure*}
	\centering
	\includegraphics[width=1\linewidth]{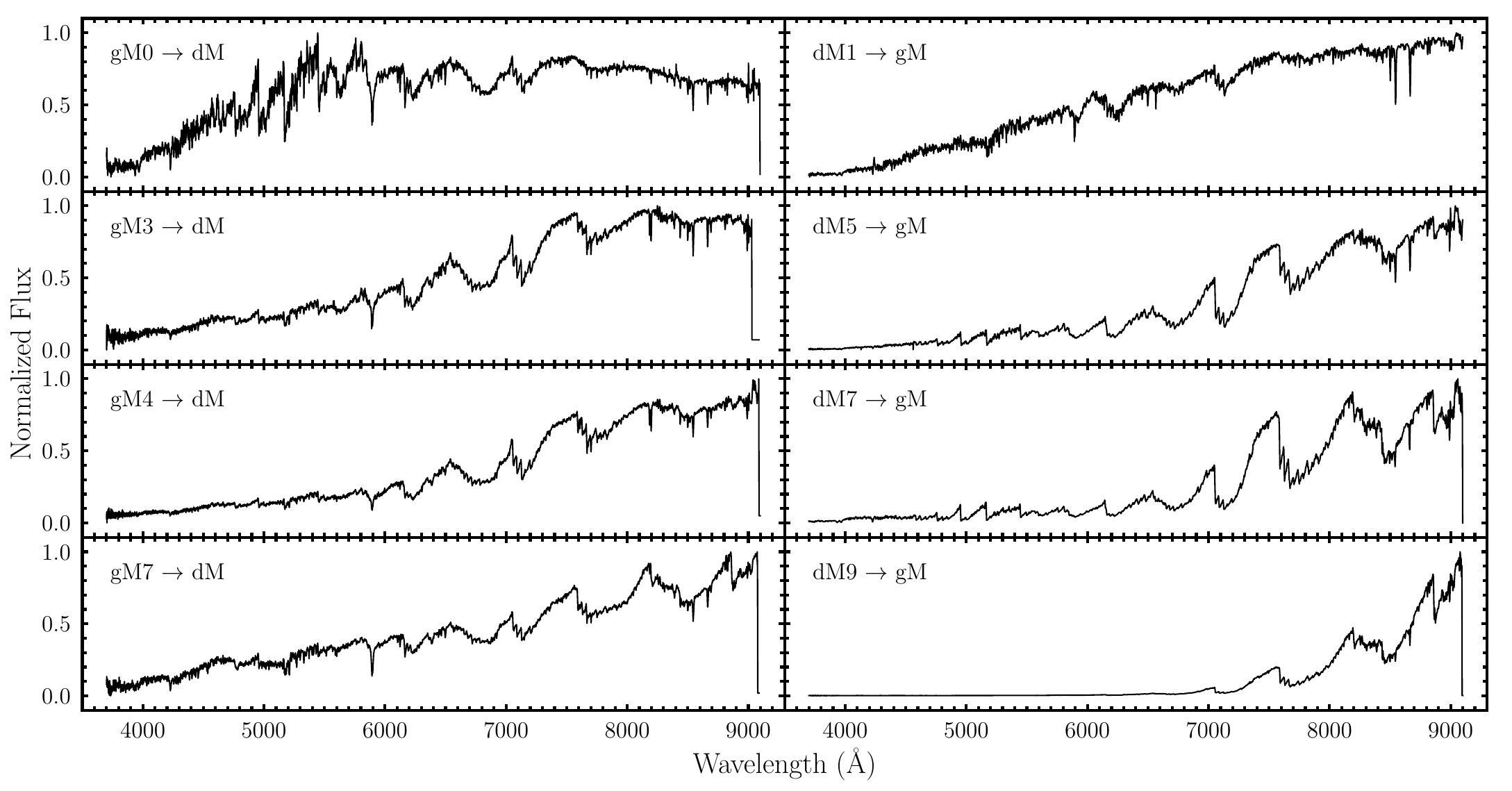}
	\caption{Examples of spectra with incorrect dM/gM classification from LAMOST. In all subfigures, the labels on the left and right of the arrows represent the luminosity classes published by LAMOST and the luminosity classes corrected in this work, respectively.}
	\label{fig:demo_spectrum_dMgM_wrong}
\end{figure*}

In the M\_Catalog\_V1, there are 52,461 spectra classified as gM by LAMOST, among which there are cases of misclassification between dM and gM. To identify as many of the misclassified dM and gM samples as possible, we  trained the CNN-dMgM model to classify all spectra in M\_Catalog\_V1 into dM and gM, with 52,225 spectra predicted as gM. The training data and labels are from M\_Catalog\_V1. In addition to removing spectra with an S/N\_max below 5, as described in Section~\ref{sec:SCM}, we excluded spectra in the dM/gM region from Figure~\ref{CMD_all_criterion} where the classification results were inconsistent with those of LAMOST to ensure the reliability of the training data and labels. From the filtered samples, we randomly selected 30,000 spectra for both dM and gM as the final training data, with the training and validation sets split in a 3:1 ratio. The confusion matrix for CNN-dMgM on the validation set is shown in Figure~\ref{fig:CM_dMgM}, with an average accuracy of 99.85\%. In Figure~\ref{CMD_all_criterion}, there are 36,202 spectra in the gM region. We combined the spectra classified as giants by LAMOST, the spectra classified as giants by the CNN-dMgM model, and the spectra in the giant star region of Figure~\ref{CMD_all_criterion}, resulting in a total of 53,987 spectra as gM candidates. The remaining spectra were considered as dM. We manually classified the 53,987 spectra into dM and gM, using the distinguishing features \citep{1991ApJS...77..417K, 2012ApJ...753...90M, 2015RAA....15.1182G} for dM and gM summarized in Table~\ref{tab:dMgM_check_feature}, and the results of the visual inspection are shown in Table~\ref{tab:dMgM_result}. Among these, LAMOST misclassified 279 dM spectra as gM, 1,799 gM spectra as dM, and some examples of the misclassified spectra by LAMOST are shown in Figure~\ref{fig:demo_spectrum_dMgM_wrong}. During the manual inspection, we discovered 85 S-type stars, 2 F-type stars, and 1,204 UNKNOWN spectra. The S-type star spectra are shown in Figure~\ref{demo_spectrum_noM}, and their identification standards follow the ZrO features mentioned in \cite{2022ApJ...931..133C}. Neither the LAMOST 1D Pipeline nor our CNN-SPT model classified S-type stars as a separate category, and they both lie in the gM region on the CMD, making it understandable that these S-type stars were found during the manual check. For the spectra directly labeled as dM, we randomly inspected about 50,000 spectra to test if any gM spectra were mixed into the dM samples. The results show that, except for 419 UNKNOWN spectra, the rest were all dM. In the previous manual check of gM and dM spectra, a total of 1,710 spectra were reclassified as non-M-type (mostly UNKNOWN), and were stored in the non-M\_Catalog\_V2. The remaining 870,518 spectra of M-type stars were stored in the M\_Catalog\_V2. With the removal of non-M spectra completed, non-M\_Catalog\_V1 and non-M\_Catalog\_V2 were combined into non-M\_Catalog, comprising 5,616 spectra in total.

\subsubsection{Radial Velocity Measurement Module}
\label{sec:RVMM}

In order to correct the LAMOST RV measurement errors as much as possible, we measured the RVs of spectra in M\_Catalog\_V2 using the Template Matching (TM) method and Doppler Shift of Absorption Lines (DSAL) method. By comparing these measurements with LAMOST RVs, we identified spectra where the LAMOST RVs might be erroneous, and we corrected the RVs through manual inspection. The specific procedure of the TM method is as follows:

1. Determining the best template (see Section~\ref{sec:spectra_templates} for a detailed description of the templates) and initial RV value. Assume that the observed spectrum's $\text{RV} \in \{ 60n \mid -10 \leq n \leq 10, n \in \mathbb{Z} \}\ (\text{km  s}^{-1})$, and calculate the $\chi^2$ value between a single template spectrum and the observed spectrum:

\begin{equation} \chi^2(\text{RV})=\frac{1}{N}\sum_{i=1}^{N} \left | O(\text{RV},\lambda_i)-P_k(\lambda_i)\times S(\lambda_i) \right |^2 \label{eq:chi2} \end{equation}

Here, $N$ is the number of wavelength points, $\lambda_i$ is the $i$-th wavelength value, $O$ is the flux of the observed spectrum, $S$ is the flux of the template spectrum, and $P_k(\lambda)$ is the $k$-th order polynomial used to reduce the errors caused by the inaccuracies in the continuum of the observed spectrum. Referring to Section~3.3 of \cite{2021ApJ...908..131Z}, we set $k=4$, and the polynomial coefficients of $P_k(\lambda)$ can be obtained by minimizing Equation (\ref{eq:chi2}). Then, we fitted the Gaussian profile to the $\chi^2(\text{RV})$ curve to determine the line center and the minimum point, which provides the RV value and the corresponding $\chi^2$ value for the single template. We identified the template with the smallest $\chi^2$ as the best template, and the RV derived from this template was taken as the initial RV value.

2. Calculating the final RV and error. After subtracting the initial RV value, we applied a velocity shift $\text{RV} \in \{ n \mid -500 \leq n \leq 500, n \in \mathbb{Z} \}\ (\text{km  s}^{-1})$ to the observed spectrum, and used the best template to measure the RV again according to Step 1. The residuals between the new RV measurement and the applied shift were then calculated. The final RV was determined as the initial RV plus the mean of the residuals, and the RV error was the standard deviation of the residuals.

\begin{table*}
	\normalsize
	\centering
	\begin{threeparttable}
	\caption{The absorption lines used in the DSAL\tnote{a} method.}

	\begin{tabular*}{\linewidth}{@{\extracolsep{\fill}}ccccccccc}
		\toprule
		\toprule
		Absorption line & K \scriptsize I & K \scriptsize I & Na \scriptsize I & Na \scriptsize I & Ti \scriptsize I & Ca \scriptsize II & Ca \scriptsize II & Ca \scriptsize II \\
		\midrule
		$\lambda$ (\AA) & 7667.0089 & 7701.0825 & 8185.5054 & 8197.0766 & 8437.2600 & 8500.3600 & 8544.4400 & 8664.5200 \\
		\bottomrule
	\end{tabular*}	
	\label{tab:absorption_line}
	\begin{tablenotes}
		\item[a]The radial velocity (RV) was calculated by measuring the Doppler shift of the absorption lines. Two DSAL methods were used in this study: DSAL I, as described in \cite{2019ApJS..240...31Z}, and DSAL II, which includes some adjustments based on DSAL I.
	\end{tablenotes}

    \end{threeparttable}
\end{table*}

This work employed two variations of the DSAL method, using eight absorption lines as shown in Table~\ref{tab:absorption_line}. The first method (DSAL I) strictly followed the procedure outlined in Section~3.2 of \cite{2019ApJS..240...31Z}. Since this method requires at least three absorption lines with consistent Doppler shifts for an accurate RV measurement (the mean and standard deviation of the RVs obtained from these absorption lines serve as the final RV and error), we designed a second method (DSAL II) to address cases where DSAL I cannot provide a result. DSAL II iteratively removed RV values that lie outside 1$\sigma$ from the measurements of the eight absorption lines. If, after $n$ iterations, the remaining RVs have a standard deviation less than 20 km s$^{-1}$ and more than one RV value, the iteration is terminated, and the mean and standard deviation of the remaining RVs are taken as the final RV and error. If fewer than two RVs remain after the $n$-th iteration, the mean and standard deviation from the $(n-1)$-th iteration are used instead. DSAL II ensures a measurement even for spectra that DSAL I cannot handle, though the precision may be lower for low-quality spectra.

\begin{figure*}
    \centering
    \includegraphics[width=1\linewidth]{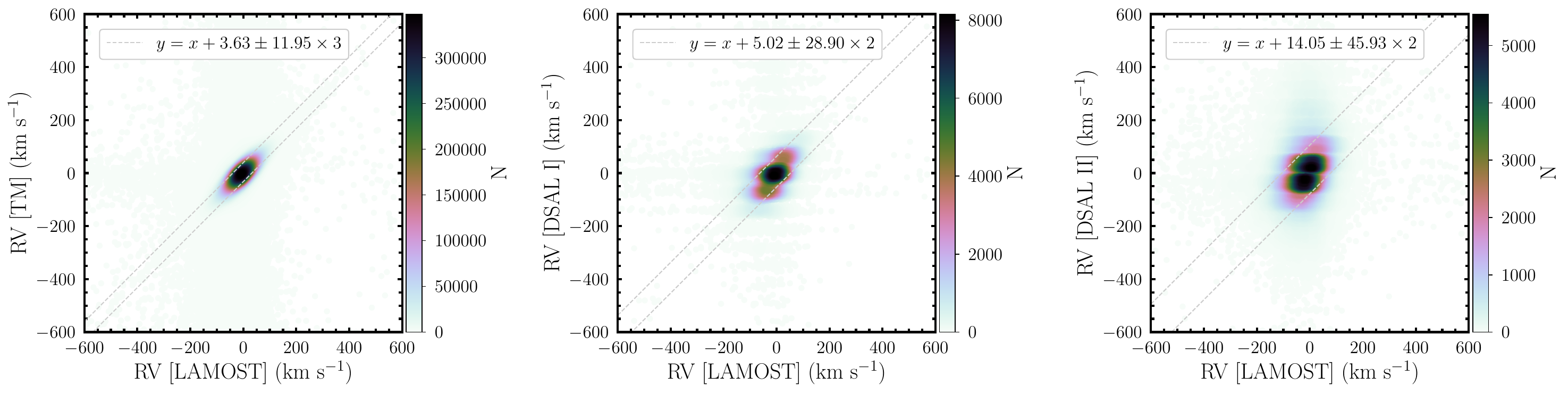}
    \caption{Comparison scatter plot of RVs derived from the Template Matching method (TM) and the two methods based on measuring the Doppler shift of absorption lines (DSAL I / DSAL II) with LAMOST RVs. In the left panel, the points within the dashed line represent those removed iteratively after a 3$\sigma$ cut, while in the middle and right panels, the points within the dashed lines correspond to those removed iteratively after a 2$\sigma$ cut. The expression for the dashed lines is provided in the upper left corner of each panel, where 3.63, 5.02, and 14.05 are the means of RV residuals, and 11.95, 28.90, and 45.93 are the standard deviations of RV residuals. The number of points outside the dashed line in the left panel corresponds to the total number of points in the middle and right panels. In the middle and right panels, the number of points outside the dashed line are 13,076 and 18,436, respectively. Together with the spectra with LAMOST RVs having absolute values greater than 150 km s$^{-1}$, a total of 34,961 spectra require manual inspection to determine the RV.}
    \label{fig:rv_compare_lamost}
\end{figure*}

Next, we outline the process of selecting spectra with potentially erroneous LAMOST RVs. First, we compare the LAMOST RVs with the results from the TM method. If they are consistent within 3$\sigma$ (the region between the two dashed lines in the left panel of Figure~\ref{fig:rv_compare_lamost}), we consider the LAMOST RV reliable. If not, we compare the LAMOST RVs with the results from the DSAL method (using DSAL I if available, otherwise using DSAL II). If the LAMOST and DSAL RV measurements are consistent within 2$\sigma$ (the region between the two dashed lines in the middle and right panels of Figure~\ref{fig:rv_compare_lamost}), the LAMOST RV is considered reliable; otherwise, we manually inspect the spectra to determine the RV. Secondly, for spectra with reliable LAMOST RVs, we also manually verify those with absolute RV values greater than 150 km s$^{-1}$ to ensure the RVs are accurate.

\begin{table*}
	\normalsize
	\centering
	\begin{threeparttable}
	\caption{The emission lines used in the DSEL\tnote{a} method.}

	\begin{tabular*}{\linewidth}{@{\extracolsep{\fill}}ccccccccc}
		\toprule
		\toprule
		Emission line & N \scriptsize II & H$\alpha$ & N \scriptsize II & S \scriptsize II  & S \scriptsize II & Ca \scriptsize II & Ca \scriptsize II & Ca \scriptsize II \\
		\midrule
		$\lambda$ (\AA) & 6549.8600 & 6564.6000 & 6585.2700 & 6718.2940 & 6732.6730 & 8500.3600 & 8544.4400 & 8664.5200 \\
		\bottomrule
	\end{tabular*}
	\label{tab:emission_line}
	\begin{tablenotes}
		\item[a]The RV was calculated by measuring the Doppler shift of the emission lines.
	\end{tablenotes}

    \end{threeparttable}
\end{table*}

\begin{table*}
	\normalsize
	\centering
	\begin{threeparttable}
	\caption{The results of the manual inspection of the RV for 34,961 spectra.}
	\begin{tabular*}{\linewidth}{@{\extracolsep{\fill}}ccccccccc}
		\toprule
		\toprule
		rv$\_$from\tnote{a} & LAMOST & TM\tnote{b} & DSAL I\tnote{c} & DSAL II\tnote{d} & TML7500\tnote{e} & DSEL\tnote{f} & NoRV\_lowQuality\tnote{g} & NoRV\_waveErr\tnote{h} \\
		\midrule
		Number & 21972 & 8278 & 995 & 3543 & 35 & 49 & 76 & 13 \\
		\bottomrule
	\end{tabular*}
    \begin{tablenotes}
		\item[a]The source of the RV, consistent with the field names in the Recommended Catalog (i.e., Table~\ref{tab:recommended_catalog}).
		\item[b]Template matching method.
		\item[c]The first DSAL method.
		\item[d]The second DSAL method.
		\item[e]Template matching method with wavelength lower than 7500 \AA.
		\item[f]The DSEL method mentioned in Table~\ref{tab:emission_line}.
		\item[g]RV could not be determined due to the low quality of the spectrum.
		\item[h]RV could not be determined due to error in the spectral wavelength calibration.
	\end{tablenotes}
	\label{tab:rv_check_reslult}
	\end{threeparttable}
\end{table*}

As shown in Figure~\ref{fig:rv_compare_lamost}, from left to right, the panels compare the RVs measured using the TM method, DSAL I method, and DSAL II method with the LAMOST RVs. In each panel, points inside the dashed lines indicate consistent RVs, while points outside the dashed lines indicate discrepancies. Ultimately, we identified 31,512 spectra from LAMOST with potentially erroneous RV measurements (corresponding to the 13,076 and 18,436 points outside the dashed lines in the middle and right panels of Figure~\ref{fig:rv_compare_lamost}). Among the spectra for which we considered the LAMOST RV reliable, there were 3,449 spectra with RV absolute values greater than 150 km s$^{-1}$. The total number of spectra requiring manual inspection, combining both groups, is 34,961. For these spectra, we moved the theoretical wavelengths of the eight absorption lines in Table~\ref{tab:absorption_line} according to the LAMOST, TM, DSAL I, and DSAL II RV measurements and marked them for further inspection, selecting a more accurate RV as the final value. During the manual inspection, two categories of spectra had unreliable RVs for all four methods. We adjusted the TM and DSAL methods and remeasured the RVs for these spectra: (1) For spectra with red-end anomalies, we remeasured their RVs using the Template Matching Method with Wavelength Limited to the Left of 7500 \AA (TML7500). (2) For spectra with emission lines, we remeasured their RVs using the Doppler Shift of Emission Lines Method (DSEL), which is analogous to DSAL II, and the eight emission lines used are listed in Table~\ref{tab:emission_line}. There were also two categories of spectra for which reliable RVs could not be determined. For these, we set the RV to null. These two categories are: (1) Spectra with insufficient quality to determine an accurate RV, and (2) Spectra with wavelength calibration errors during LAMOST spectral data processing caused abnormal absorption line positions. The RV inspection results for the 34,961 spectra are summarized in Table~\ref{tab:rv_check_reslult}, and the corrected RVs were stored in M\_Catalog\_V3. Columns~2--7 of Table~\ref{tab:rv_check_reslult} show the sources and quantities of manually inspected RVs, among which Columns~3--7 correspond to the updated RVs, totaling 12,900. The last two columns show the numbers of spectra for which reliable RVs could not be provided due to the two specific issues, totaling 89.

It should be noted that, for spectra without manual inspection, we adopted the LAMOST RV as the final value. Among these spectra, 3.56\% show that the LAMOST RV is consistent with TM but not with DSAL, whereas 7.59\% show that the LAMOST RV is consistent with DSAL but not with TM. These two cases highlight potential risks in our RV determinations: (1) the LAMOST and TM RVs are consistent with each other but may nevertheless deviate from the true value; (2) the same issue may arise for the LAMOST and DSAL RVs. Even under the most unfavorable assumption, however, only about 11\% of the RV results would exhibit relatively large errors.

\begin{figure*}
	\centering
	\includegraphics[width=1\linewidth]{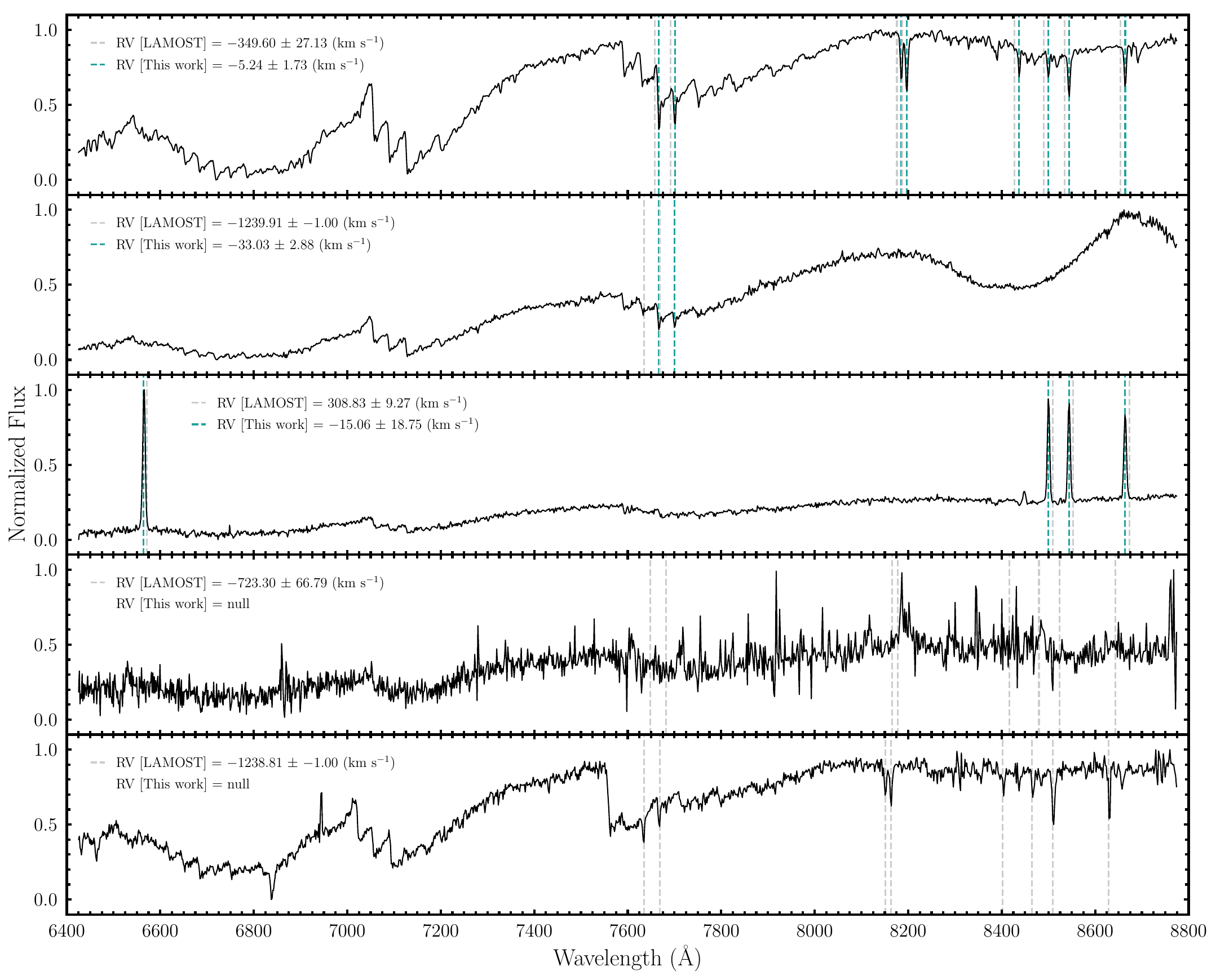}
	\caption{Examples of spectra with incorrect RV measurements from LAMOST. The LAMOST RV and those corrected in this work are given in all subplots. The top three panels show normal spectra, red-end abnormal spectra, and spectra with emission lines where the LAMOST RV is incorrectly measured. The bottom two panels show spectral examples with incorrect RV measurements of LAMOST, for which RV values cannot be provided in this work. These issues are due to poor spectral quality and wavelength calibration errors, respectively.}
	\label{fig:rv_wrong_demo}
\end{figure*}

Figure~\ref{fig:rv_wrong_demo} shows examples of spectra with corrected RVs. Subplots~1--3 display the spectra with normal, red-end anomaly, and emission line features. The red-end anomaly occurs when incandescent lamps were inadvertently left on during observations, contaminating the spectra with spurious features\footnote{After completing the entire workflow (see Figure~\ref{workflow_diagram}), we compiled statistics for the 35 red-end-anomalous spectra, all of which were visually confirmed to be M dwarfs. Our statistic results are as follows: CNN-SPT and CNN-WDM classified 19 and 18 spectra as M, respectively; for those spectra that at least one of these two models labeled as non-M, visual inspection confirmed them to be M-type stars; CNN-dMgM classified 26 spectra as dM, and the remaining 9 spectra were confirmed as M dwarfs by visual inspection; and for CNN-TLMA, only three spectra had predicted parameters that passed the quality-assessment cuts and were recommended for use. Given the extensive red-end anomalies in these spectra, relatively lower classification and parameter-prediction accuracies for this subset are to be expected.}. Subplots~4 and 5 display spectra for which accurate RVs could not be determined, corresponding to low-quality spectra and those with wavelength calibration errors. The gray and green dashed vertical lines represent the wavelengths of absorption or emission lines shifted according to the LAMOST RV or the corrected RV, respectively, and the RV values are annotated in the top left corner. As shown in Subplots~1--3, the corrected RVs are more reliable than the original LAMOST RVs. Notably, in Subplot~5, the RV of -1238.81 km~s$^{-1}$ measured by LAMOST appears reasonable based on the Doppler shift of the absorption lines. However, for stars, a line-of-sight velocity greater than 1000 km~s$^{-1}$ is unusually high, and we need to confirm whether they are hypervelocity stars or if there is an issue with spectral processing. After inspection by LAMOST official personnel, a wavelength calibration error was confirmed in this spectrum.

\subsubsection{Stellar Parameter Measurement Module}
\label{sec:SPMM}

In this work, we used the ``label transfer + parameter prediction" strategy for measuring parameters of M-type stars. In this work, ``label transfer" refers to the approach of using LAMOST low-resolution M-type stellar spectra and high-precision M-type stellar parameters measured from APOGEE high-resolution spectra to train deep learning models. After the model learns the relationship between low-resolution spectral features and stellar parameters, this knowledge will be subsequently ``transferred" to other LAMOST low-resolution M-type stellar spectra, enabling more reliable stellar parameter measurements. Compared to directly applying models trained on synthetic spectra to observed spectra for stellar parameter prediction, the ``label transfer'' approach significantly reduces systematic errors arising from the ``synthetic gap'' \citep{2020A&A...642A..22P}, which is particularly advantageous given that current atmospheric models for M-type stars remain relatively immature, with significant discrepancies between theoretical and observed spectra \citep{2023MNRAS.523.1297J}. Furthermore, the ``parameter prediction" methodology—directly estimating stellar parameters from spectra—generally has higher efficiency compared to ``parameter inference" approaches.

We used the CNN-TLMA model (The network architecture is shown in Figure~\ref{CNN_framework}) to measure four stellar parameters ($T_\text{eff}$, $\text{log}\ g$, [M/H], and [$\alpha$/M]) for all spectra in M\_Catalog\_V3 and provided the associated errors. The measurement results were subsequently compiled into the Recommended Catalog. It is important to note that, as described in Section~\ref{sec:preprocess}, the input to CNN-TLMA is the spectrum shifted to the rest frame using the updated RVs from RVMM (see Section~\ref{sec:RVMM} for more details), and the RV shift is applied only for this module. The training data of the CNN-TLMA is from M\_Catalog\_V3, and the labels are the four stellar parameters provided by APOGEE DR16 (hereafter, unless otherwise specified, referred to as APOGEE). Beyond removing spectra with an S/N\_max below 5, as described in Section~\ref{sec:SCM}, we further ensured that the \texttt{ASPCAP\_CLASS}\footnote{The \texttt{ASPCAP\_CLASS} is a field provided in the APOGEE DR16 parameter catalog, representing the temperature class of the best-fitting spectrum.} was consistent with the DGCM (see Section~\ref{sec:DGCM} for more details) classification results to ensure the reliability of the training data and labels. All the filtered samples were used as the final training data, with 12,541 dM spectra and 3,927 gM spectra. We trained 10 CNN-TLMA models for the dM and gM spectra using ten-fold cross-validation, where the training data was randomly divided into 10 equal-sized subsets. In each fold, one subset was selected as the validation set, and the remaining 9 subsets were used as the training set to train the model. The mean and standard deviation of the predictions from the 10 models were used as the final stellar parameters and their corresponding errors.

\begin{table*}
	\normalsize
	\centering
	\caption{Residual statistics and loss values for the CNN-TLMA models of dM and gM stars on the validation set in ten-fold cross-validation.}
	\centering
	\begin{tabular*}{\textwidth}{@{\extracolsep{\fill}}cccccc}
		\toprule
		\toprule
		\multicolumn{6}{c}{dM} \\
		\midrule
		Model ID & $\Delta\ T_\text{eff}$ (K) & $\Delta$ log $g$ (dex) & $\Delta$ [M/H] (dex) & $\Delta$ [$\alpha$/M] (dex) & Validation Loss \\
		\midrule
		0 & $3 \pm 50$	& $-0.00 \pm 0.15$	& $-0.01 \pm 0.13$	& $0.00 \pm 0.04$	& 0.282\\
		1 & $3 \pm 51$	& $0.04 \pm 0.16$	& $0.01 \pm 0.14$	& $-0.00 \pm 0.04$ 	& 0.323\\
		2 & $-17 \pm 52$& $-0.01 \pm 0.17$	& $-0.01 \pm 0.14$ 	& $0.00 \pm 0.04$ 	& 0.322\\
		3 & $-4 \pm 47$	& $-0.01 \pm 0.15$	& $-0.02 \pm 0.13$ 	& $0.00 \pm 0.04$ 	& 0.278\\
		4 & $-3 \pm 51$	& $0.02 \pm 0.16$	& $0.00 \pm 0.13$ 	& $-0.00 \pm 0.04$ 	& 0.300\\
		5 & $7 \pm 52$	& $0.00 \pm 0.16$	& $-0.00 \pm 0.14$ 	& $-0.00 \pm 0.04$ 	& 0.304\\
		6 & $-5 \pm 51$	& $0.02 \pm 0.17$	& $0.01 \pm 0.13$ 	& $0.01 \pm 0.04$ 	& 0.295\\
		7 & $3 \pm 54$	& $0.01 \pm 0.16$	& $0.03 \pm 0.14$ 	& $0.01 \pm 0.04$ 	& 0.302\\
		8 & $5 \pm 55$	& $-0.02 \pm 0.17$	& $0.02 \pm 0.15$ 	& $-0.00 \pm 0.05$ 	& 0.364\\
		9 & $6 \pm 52$	& $0.00 \pm 0.16$	& $0.03 \pm 0.13$ 	& $-0.00 \pm 0.04$ 	& 0.283\\
		\midrule
		\multicolumn{6}{c}{gM} \\
		\midrule
		0 & $-1 \pm 37$ & $-0.01 \pm 0.18$ 	& $-0.00 \pm 0.10$ 	& $-0.00 \pm 0.04$ 	& 0.185\\
		1 & $-1 \pm 29$ & $-0.01 \pm 0.14$ 	& $-0.00 \pm 0.10$ 	& $0.00 \pm 0.04$ 	& 0.137\\
		2 & $9 \pm 30$ 	& $-0.00 \pm 0.14$ 	& $-0.01 \pm 0.11$ 	& $0.00 \pm 0.05$ 	& 0.161\\
		3 & $-7 \pm 32$ & $-0.03 \pm 0.14$ 	& $0.00 \pm 0.09$ 	& $0.00 \pm 0.04$ 	& 0.153\\
		4 & $3 \pm 40$ 	& $0.00 \pm 0.16$ 	& $-0.02 \pm 0.11$ 	& $0.00 \pm 0.05$ 	& 0.212\\
		5 & $1 \pm 28$ 	& $0.00 \pm 0.13$ 	& $0.01 \pm 0.10$ 	& $0.01 \pm 0.05$ 	& 0.159\\
		6 & $-1 \pm 33$ & $0.01 \pm 0.15$ 	& $0.01 \pm 0.10$ 	& $0.00 \pm 0.04$ 	& 0.146\\
		7 & $-5 \pm 35$ & $0.03 \pm 0.14$ 	& $0.01 \pm 0.10$ 	& $0.00 \pm 0.04$ 	& 0.160\\
		8 & $-0 \pm 29$ & $0.01 \pm 0.15$ 	& $0.01 \pm 0.12$ 	& $-0.00 \pm 0.04$ 	& 0.168\\
		9 & $9 \pm 37$ 	& $0.00 \pm 0.21$ 	& $-0.00 \pm 0.11$ 	& $0.01 \pm 0.04$ 	& 0.205\\
		\bottomrule
	\end{tabular*}
	
	\label{tab:valid_loss}
\end{table*}

\begin{figure*}
	\centering
	\includegraphics[width=1\linewidth]{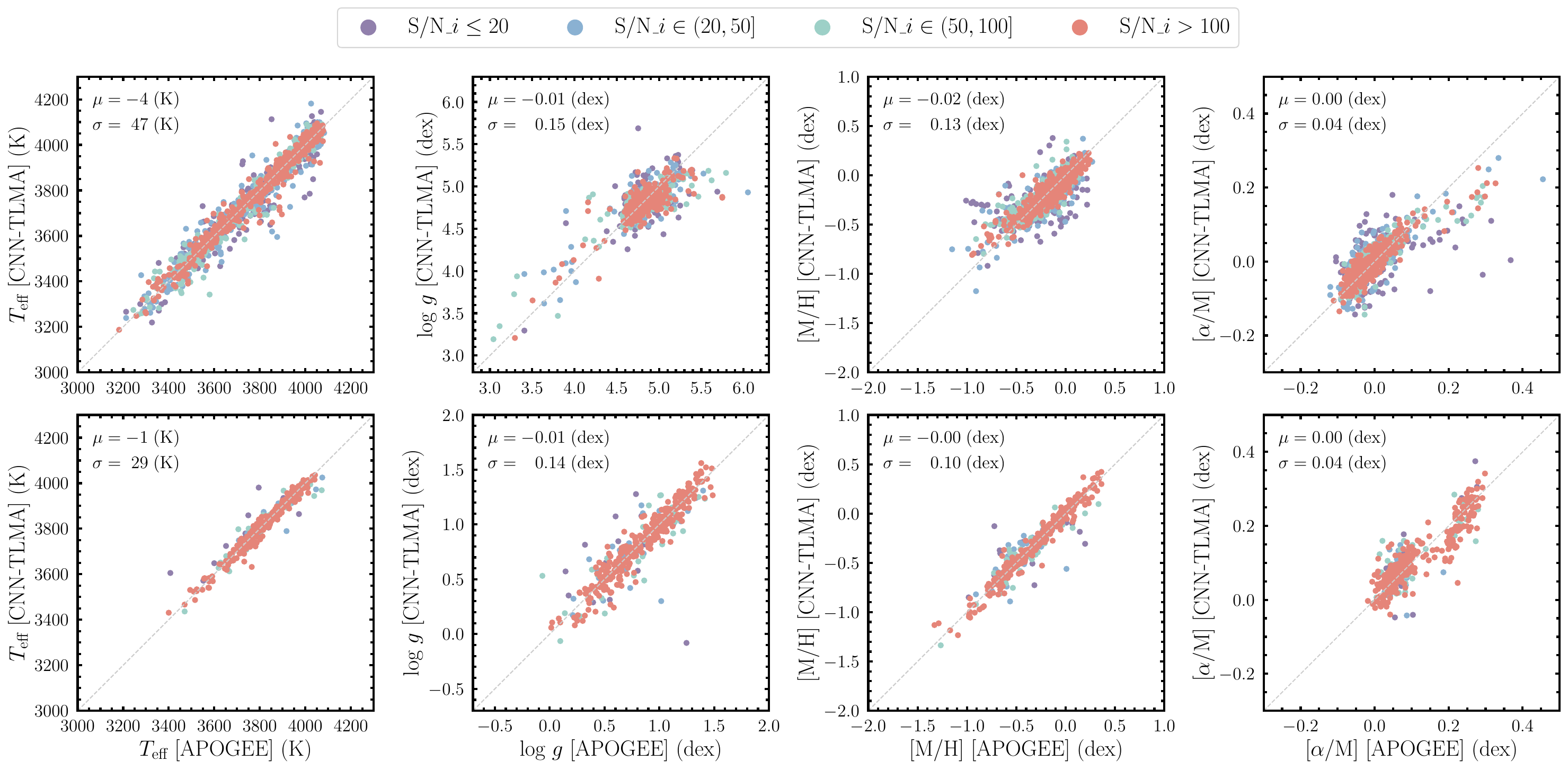}
	\caption{Comparison of the CNN-TLMA model predictions (with the smallest validation loss in ten-fold cross-validation) and APOGEE data. The first row presents results for dM, while the second row shows results for gM. Spectra in different signal-to-noise ratio ranges are represented by scatter points in distinct colors.}
	\label{fig:loss_min}
\end{figure*}

The residual statistics and loss values for these 20 models on their respective validation sets are summarized in Table~\ref{tab:valid_loss}. It can be seen that the performance differences on the validation set across the folds are minimal, so we only show the comparison of the four stellar parameters predicted by the model with the minimum validation loss in Figure~\ref{fig:loss_min}. It can be seen that there is a good consistency between the predicted values and the true labels, with outliers mostly being low S/N spectra. The mean predictions from multiple models smooth out the bias of individual models, improving the robustness of the predictions, while the standard deviation effectively quantifies the uncertainty between models, providing confidence in the results. Furthermore, the ten-fold cross-validation fully utilizes the data, mitigates the risk of overfitting, and enhances the model's generalization ability and the reliability of the results. Therefore, it is reasonable to use model ensemble, taking the mean of the predictions from the 10 models as the final result, with the standard deviation as the error quantification metric. For the training data, the integrated predicted results (including \(T_\text{eff}\), \(\log g\), [M/H], [\(\alpha\)/M]) show the systematics and dispersions relative to APOGEE as follows: (1) for dM stars: \(0 \pm 52\) K, \(0.00 \pm 0.16\) dex, \(0.01 \pm 0.14\) dex, and \(0.00 \pm 0.04\) dex; (2) for gM stars: \(1 \pm 34\) K, \(0.00 \pm 0.16\) dex, \(0.00 \pm 0.10\) dex, and \(0.00 \pm 0.04\) dex.

\section{Result}
\label{sec:Result}

\subsection{Quality Assessment Cuts and Recommended Catalog}
\label{sec:quality_assessment}

As mentioned earlier, RVMM (described in detail in Section~\ref{sec:RVMM}) corrected the erroneous RVs from LAMOST. SPMM (described in detail in Section~\ref{sec:SPMM}) estimated the stellar parameters and their errors ($T_\text{eff}$, log $g$, [M/H] and [$\alpha$/M]) separately for dM and gM spectra. These parameters are included in the Recommended Catalog (a summary of the field descriptions is presented in Table~\ref{tab:recommended_catalog}) provided in this work, which also contains quality assessment cut flags for RV (\texttt{rv\_flag}) and stellar parameters (\texttt{parameter\_flag}). Further details on these flags are provided below.

1. Before January 14, 2012, some exposures in the LAMOST low-resolution survey were affected by issues with the wavelength calibration lamp in the red channel, which could have led to errors in RVs. To ensure data reliability, we applied a selection criterion to the \texttt{rv} values in the Recommended Catalog. The \texttt{rv\_flag} is set to 1 if the following conditions are met (indicating that the RV is recommended for use); otherwise, it is set to 0 (indicating that the RV is either unavailable or not recommended for use).

\begin{enumerate}[label=(\arabic*)]
    \item \texttt{rv} is not null.
    \item The spectrum was observed after January 14, 2012.
\end{enumerate}

2. We applied quality assessment cuts to the four stellar parameters in the Recommended Catalog. For stellar parameters recommended for use, the \texttt{parameter\_flag} is set to 1; otherwise, it is set to 0. Since the primary spectral features used in this work are concentrated around the $i$ band, the accuracy of the predicted parameters is likely to be affected if the $i$ band spectrum has quality issues. Therefore, for spectra with S/N\_$i$ (S/N in $i$ band) less than or equal to 0, the \texttt{parameter\_flag} is directly set to 0. For spectra with S/N\_$i$ greater than 0, the \texttt{parameter\_flag} will be set through the following two steps:

\begin{enumerate}[label=(\arabic*)]
    \item The errors of the stellar parameters are within a reasonable range.
    \label{step:parameter_error_range}
    
    For spectra with the same S/N level, if the four stellar parameters predicted by SPMM are similar, their error levels should also be comparable. We divided the dM and gM samples into four groups based on S/N\_$i$, corresponding to the following S/N\_$i$ ranges: (0, 20], (20, 50], (50, 100], and $>$ 100. For samples in different S/N\_$i$ intervals, we used Linear Regression (LR) to fit the relationship between stellar parameter errors and both the stellar parameters and S/N\_$i$. The specific formula is given below:
    
    \begin{multline}
    \hat{\sigma}_p = a_0 + a_1\cdot\text{S/N}\_i + a_2\cdot T_\text{eff} + a_3\cdot\log g \\
    + a_4\cdot\text{[M/H]} + a_5\cdot[\alpha/\text{M}]
    \label{eq:LR}
    \end{multline}
    
    Here, $\hat{\sigma}_p$ represents the LR-predicted value of the stellar parameter error $\sigma_p$, where $p$ corresponds to $T_\text{eff}$, log $g$, [M/H], or [$\alpha$/M]. We define the reasonable range of parameter errors as the region where the difference between the actual error and the LR-predicted error ($\sigma_p - \hat{\sigma}_p$) satisfies the following condition:
    
    \begin{equation}
    \sigma_p-\hat{\sigma}_p<\text{Mean}(\sigma_p-\hat{\sigma}_p)+3\cdot \text{STD}(\sigma_p-\hat{\sigma}_p)
    \label{eq:parameter_error_flag}
    \end{equation}
    
    Here, Mean($\cdot$) represents the mean of $\sigma_p - \hat{\sigma}_p$, and STD($\cdot$) denotes its standard deviation. A smaller $\sigma_p - \hat{\sigma}_p$ is preferred, while excessively large values indicate reduced reliability of the stellar parameters. Therefore, the \texttt{parameter\_flag} is initially set to 1 for stellar parameters that satisfy Equation~(\ref{eq:parameter_error_flag}), and to 0 otherwise.
    \item The stellar parameters are within a reasonable range.
    
    We first identify reliable members with trustworthy stellar parameters from the training data used by SPMM based on the following criteria:
    
    \begin{enumerate}[label=(\alph*)]
        \item \texttt{parameter\_flag} = 1
        \item \texttt{snr\_u}, \texttt{snr\_g}, \texttt{snr\_r} and \texttt{snr\_z} $>$ 0
        \item \texttt{snr\_i} $>$ 20
        \item \texttt{dMgM}, \texttt{ASPCAP\_CLASS} and \texttt{CMD\_pos} agree on the dM/gM classification
    \end{enumerate}
    
    Here, \texttt{snr\_u/g/r/i/z}, \texttt{dMgM} and \texttt{CMD\_pos} are fields provided in the Recommended Catalog (see Table~\ref{tab:recommended_catalog}). For the description of \texttt{ASPCAP\_CLASS} refer to the footnote in Section~\ref{sec:SPMM}. Based on this, the reasonable ranges of the four stellar parameters are obtained, as shown in Table~\ref{tab:para_range}. Subsequently, we apply the reasonable parameter ranges to the samples whose \texttt{parameter\_flag} was initially set to 1 in step~\ref{step:parameter_error_range}, reassigning \texttt{parameter\_flag} to 0 for those falling outside the ranges.
\end{enumerate}

\begin{figure}
	\centering
	\includegraphics[width=1\linewidth]{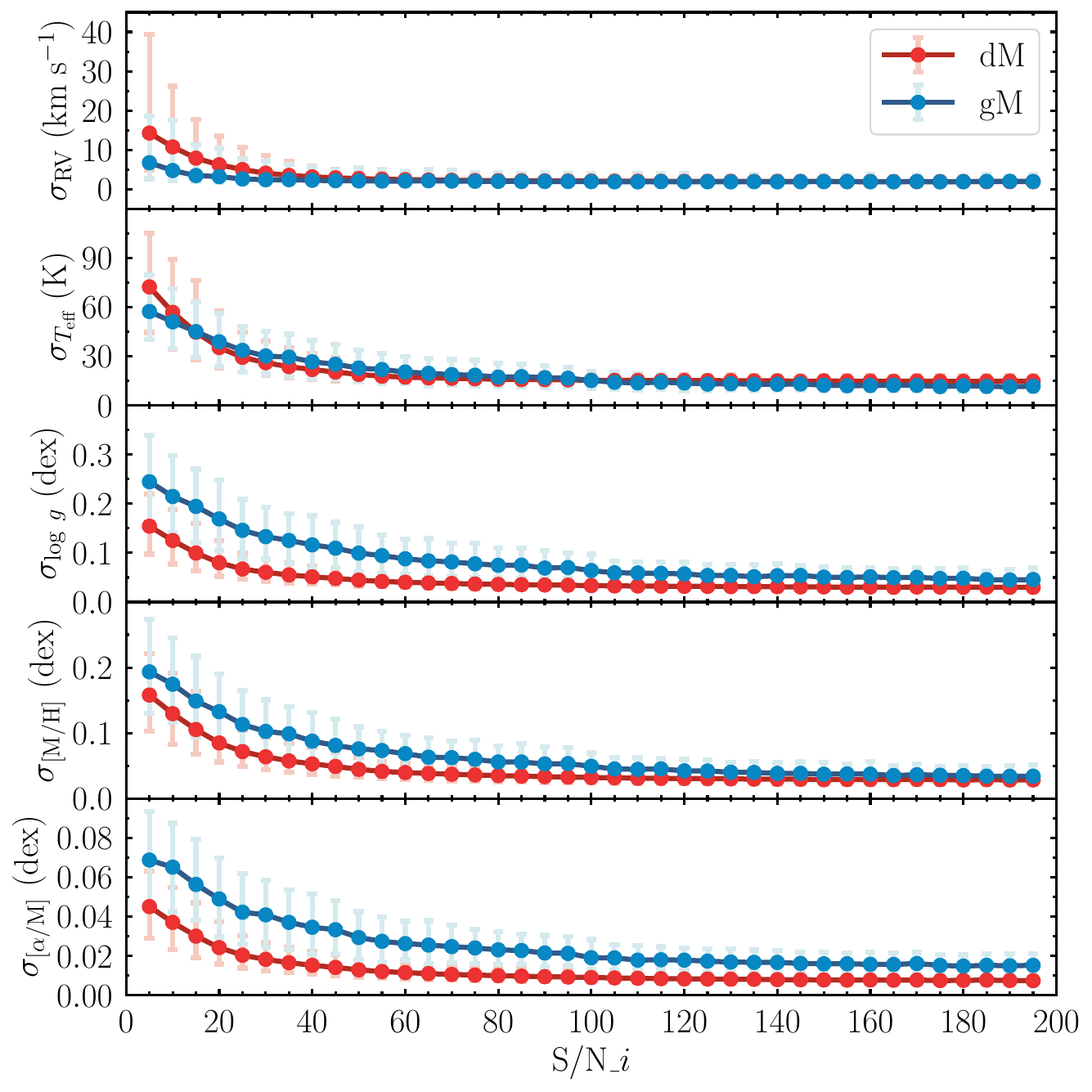}
	\caption{Errors of the recommended RVs and stellar parameters as a function of S/N\_$i$. Each point represents the median value within the interval (S/N\_$i-2.5$, S/N\_$i+2.5$], and the error bars indicate the 16th and 84th percentiles in each bin. Spectra with S/N\_$i\ge200$ are not shown, as the associated errors remain nearly constant in this regime.}
	\label{fig:para_vs_snr}
\end{figure}

When using the RVs provided in the Recommended Catalog of this work, we suggest selecting those with \texttt{rv\_flag}$=$1 (94\% of the total spectra). For stellar parameters, we recommend using those with \texttt{parameter\_flag}$=$1, corresponding to 716,207 dM spectra (87\% of the total dM spectra) and 40,907 gM spectra (82\% of the total gM spectra). Figure~\ref{fig:para_vs_snr} displays the internal error levels of the recommended RVs and stellar parameters across different S/N\_$i$ intervals, where red corresponds to dM and blue corresponds to gM. As shown, the errors gradually decrease and tend to stabilize with increasing S/N\_$i$. The average internal errors for dM/gM are respectively: RV 6/3 km s$^{-1}$, $T_\text{eff}$ 30/17 K, log $g$ 0.07/0.07 dex, [M/H] 0.07/0.05 dex, and [$\alpha$/M] 0.02/0.02 dex.

\begin{table*}
	\normalsize
	\centering
	\setlength{\tabcolsep}{3pt}
    \begin{threeparttable}
        \caption{
        Recommended Catalog.
        }
    	\begin{tabular*}{\linewidth}{@{\extracolsep{\fill}}cc>{\centering\arraybackslash}p{0.7\linewidth}}
    		\toprule
    		\toprule
    		Field & Unit & Description \\
    		\midrule
    		obsid & ... & The unique spectral ID from LAMOST DR10 \\
    		ra & degree & Right ascension from LAMOST DR10 \\
    		dec & degree & Declination from LAMOST DR10 \\
    		snr\_u/g/r/i/z & ... & The signal-to-noise ratio in the $u$, $g$, $r$, $i$ or $z$ band \\
    		orp\_u/g/r/i/z & ... & The proportion of ormask values that are non-zero in the $u$, $g$, $r$, $i$ or $z$ band \\
    		subclass & ... & The stellar spectral type from LAMOST DR10 \\
    		luminosity\_class & ... & The corrected dM/gM classification result \\
    		rv & km s$^{-1}$ & The corrected RV \\
    		rv\_err & km s$^{-1}$ & The error of RV \\
    		rv\_from & ... & The source of RV \\
    		teff & K & Effective temperature \\
    		teff\_err & K & The error of the effective temperature \\
    		logg & dex & Surface gravity \\
    		logg\_err & dex & The error of the surface gravity \\
    		m\_h & dex & Metallicity \\
    		m\_h\_err & dex & The error of the Metallicity \\
    		alpha\_m & dex & Alpha element abundance \\
    		alpha\_m\_err & dex & The error of the alpha element abundance \\
    		rv\_flag\tnote{a} & ... & The RV validity flag\\
    		parameter\_flag\tnote{b} & ... & The stellar parameter quality flag \\
    		source\_id & ... & The unique source identifier from Gaia DR3 \\
    		r\_med\_geo & pc & The median of the geometric distance posterior from \cite{2021AJ....161..147B}\\
    		ebv\tnote{c} & mag & The color excess $\text{E}(B-V)$ \\ 
    		bp\_rp\_0\tnote{d}& mag & The extinction-corrected Gaia $G_\text{BP}-G_\text{RP}$ colour \\
    		M\_G\_0& mag & The absolute magnitude $M_\text{G}$ in Gaia's $G$-band, calculated using Equation (\ref{eq:M_G}) \\
    	    CMD\_flag\tnote{e} & ... & The Gaia astrometric parameter quality flag\\
    		CMD\_pos\tnote{f}& ... & The position of sources with CMD\_flag=1 on the Gaia color-magnitude diagram \\
    		W1\_W2\_0\tnote{g} & mag & The extinction-corrected AllWISE \citep{2014yCat.2328....0C} $W1-W2$ colour \\
    		e\_W1/W2 & mag & The mean error of AllWISE $W1/W2$ magnitude\\
    		\bottomrule
    	\end{tabular*}
        \label{tab:recommended_catalog}
    	\begin{tablenotes}
    	    \item[a] \begin{CJK*}{UTF8}{gbsn}rv\_flag = 1 indicates RV is recommended for use, 0 indicates RV is unavailable or there may be wavelength calibration issues in the spectrum.\end{CJK*}
    		\item[b] \begin{CJK*}{UTF8}{gbsn}parameter\_flag = 1 indicates that the stellar parameters are recommended for use, while 0 indicates they are not recommended for use.\end{CJK*}
    		\item[c] \begin{CJK*}{UTF8}{gbsn}$\text{E}(B-V)$ is determined from the 3D dust map Bayestar19 \citep{2019ApJ...887...93G} using r\_med\_geo\end{CJK*}
    		\item[d] \begin{CJK*}{UTF8}{gbsn}The extinction coefficient is from \cite{2018MNRAS.479L.102C}.\end{CJK*}
    		\item[e] \begin{CJK*}{UTF8}{gbsn}Based on the five equations in Section~\ref{sec:SCM}, the Gaia astrometric data are quality-filtered. CMD\_flag = 1 indicates that the Gaia astrometric data are reliable, while 0 indicates they are unreliable.\end{CJK*}
    		\item[f] \begin{CJK*}{UTF8}{gbsn}CMD\_pos = `dM' indicates the star is in the dM region of Figure~\ref{CMD_all_criterion}, `gM' in the gM region, and `other' in the non-M region.\end{CJK*}
    		\item[g] \begin{CJK*}{UTF8}{gbsn}The extinction coefficient is from \cite{2023ApJS..264...14Z}.\end{CJK*}

    	\end{tablenotes}
    \end{threeparttable}
\end{table*}

\begin{table}
	\normalsize
	\centering
    \begin{threeparttable}
        \caption{Recommended range of stellar parameters.}
    	\begin{tabular*}{\linewidth}{@{\extracolsep{\fill}}ccc}
    		\toprule
    		\toprule
    		Stellar parameter & dM & gM \\
    		\midrule
    		$T_\text{eff}$ (K) & $[3225,4125]$ & $[3425,4075]$  \\

    		log $g$ (dex) & $[4.35,5.45]$ & $[-0.05,1.55]$  \\

    		$\text{[M/H]}$ (dex) & $[-0.85,0.35]$ & $[-1.35,0.45]$  \\

    		$[\alpha/\text{M}]$ (dex) & $[-0.13,0.25]$ & $[-0.03,0.31]$  \\
    		\bottomrule
    	\end{tabular*}
    	\label{tab:para_range}
    \end{threeparttable}
\end{table}

\subsection{External Comparison}
\label{sec:external_comparison}

To facilitate subsequent analysis, before performing external comparisons, we define the following spectral quality criteria to ensure, to the greatest extent possible, the use of reliable parameters from both our work and the literature:
\begin{enumerate}[label=(\arabic*)]
    \item \texttt{snr\_u} and \texttt{snr\_g} $>$ 0
    \item \texttt{snr\_r}, \texttt{snr\_i}, and \texttt{snr\_z} $>$ 20
    \item \texttt{orp\_u}, \texttt{orp\_g}, \texttt{orp\_r}, \texttt{orp\_i}, and \texttt{orp\_z} $<$ 1\%
\end{enumerate}

Here, \texttt{orp\_u/g/r/i/z} are fields provided in the recommended catalog (see Table~\ref{tab:recommended_catalog}). 

\subsubsection{RV}

To assess the reliability of the RVs recommended in this work (corresponding to \texttt{rv\_flag}$=$1 in Table~\ref{tab:recommended_catalog}), we cross-matched with the RVs provided in Gaia DR3 \citep{2023A&A...674A...5K}. We only consider sources that meet the following criteria to maximize the use of reliable Gaia RVs:

\begin{enumerate}[label=(\arabic*)]
    \item \texttt{radial\_velocity\_error} $<$ 2 km s$^{-1}$
    \item \texttt{rv\_expected\_sig\_to\_noise} $>$ 20
\end{enumerate}

Here, both of these fields are provided in the Gaia catalog. \texttt{radial\_velocity\_error} is the combined RV formal uncertainty, while \texttt{rv\_expected\_sig\_to\_noise} is the expected S/N in the combination of the spectra used to obtain the RV.

\begin{figure}
	\centering
	\includegraphics[width=1\linewidth]{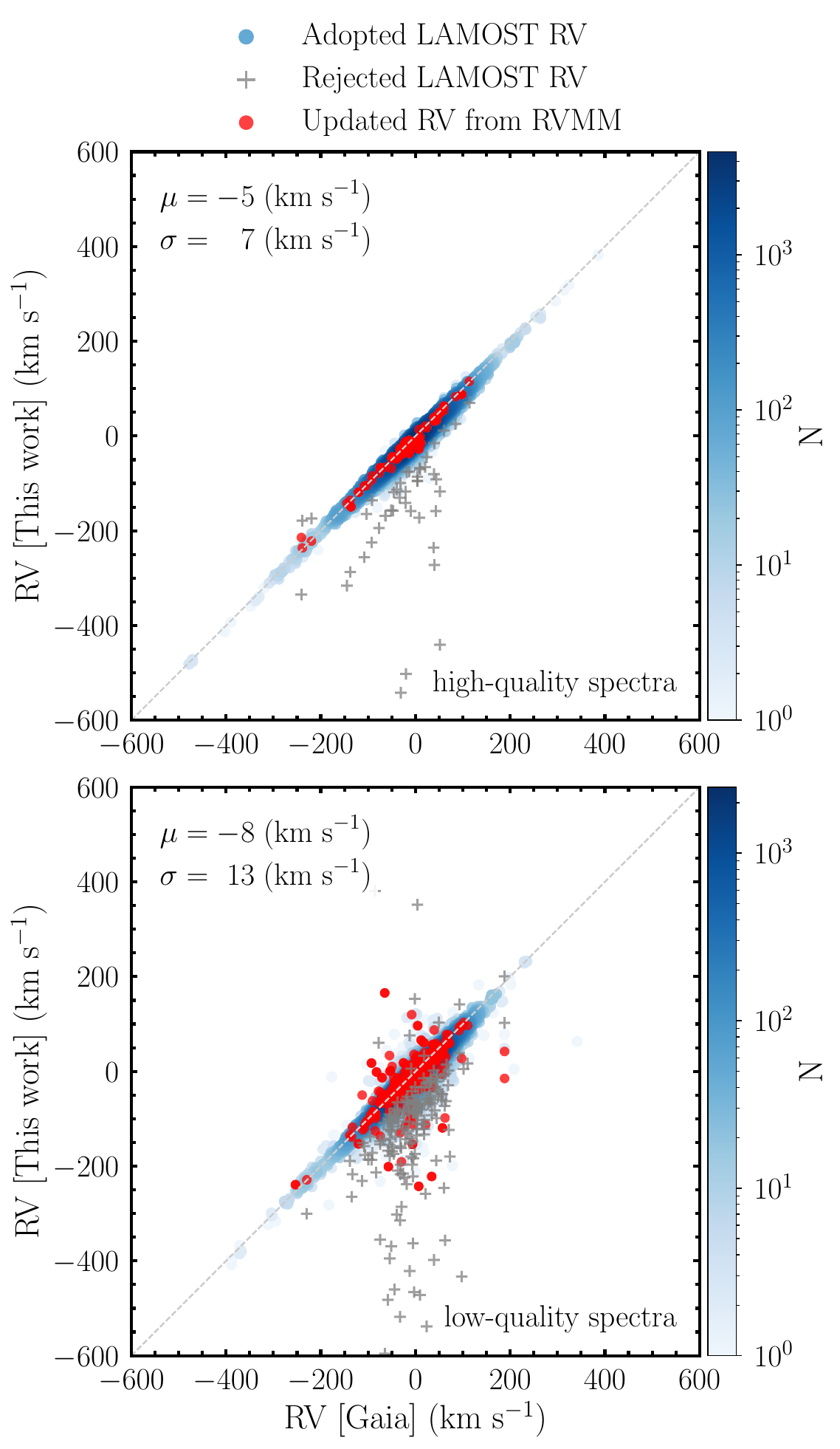}
	\caption{Comparison between the recommended RVs and the Gaia RVs. The upper and lower panels correspond to high-quality and low-quality spectra, respectively. The blue points and gray plus signs represent LAMOST RVs that were deemed reliable and unreliable by RVMM (described in detail in Section~\ref{sec:RVMM}), respectively. The red points show the corresponding updated RVs from RVMM, with each red point paired with its corresponding gray plus sign. The $\mu$ and $\sigma$ in each panel represent the systematic difference and dispersion between the recommended RVs and the Gaia RVs, respectively.}
	\label{fig:rv_vs_gaia}
\end{figure}

Figure~\ref{fig:rv_vs_gaia} presents a comparison between the RVs provided in this work and those from Gaia. The upper panel shows high-quality spectra meeting the criteria defined in Section~\ref{sec:external_comparison}, while the lower panel shows low-quality spectra failing to satisfy the criteria. In the figure, the y-coordinates of blue points represent LAMOST RVs that were deemed reliable by RVMM (described in detail in Section~\ref{sec:RVMM}), while gray plus signs (some of which extend beyond the plot boundaries and are not displayed) indicate LAMOST RVs that were deemed unreliable by RVMM, which were then updated by RVMM and shown as red points. As shown in Figure~\ref{fig:rv_vs_gaia}, for high-quality spectra (upper panel), the RVs recommended in this work are highly consistent with Gaia RVs, accurately correcting the measurement errors in the original LAMOST RVs. For low-quality spectra (lower panel), although the dispersion between the recommended RVs and Gaia RVs increases due to noise features sometimes being misidentified as absorption lines, a substantial improvement in accuracy over the original LAMOST RVs is still achieved.

\subsubsection{Stellar Parameters}
\label{sec:external_comparision_parameter}

To assess the reliability of the stellar parameters recommended in this work (corresponding to \texttt{parameter\_flag}$=$1 in Table~\ref{tab:recommended_catalog}), we conducted comparisons not only with the parameters provided by the LASPM \citep{2021RAA....21..202D}, but also with those from \citet[hereafter Li2021]{2021ApJS..253...45L}, \citet[hereafter Ding2022]{2022ApJS..260...45D}, \citet[hereafter Qiu2023]{2023RAA....23e5008Q}, and \citet[hereafter Liang2022]{2022AJ....163..153L}.

LASPM used a template-matching algorithm and BT-Settl theoretical spectra to measure $T_\text{eff}$, log $g$, and [M/H] for all M-type stars observed by LAMOST. Ding2022 employed the ULySS software package and template spectra generated with the MILES interpolator to determine $T_\text{eff}$, log $g$, and [M/H] for a total of 763,136 spectra in LAMOST DR8. Li2021 applied the Stellar Label Machine (SLAM) method, with stellar parameter labels from APOGEE DR16, to measure $T_\text{eff}$ and [M/H] for 243,231 selected dM spectra in LAMOST DR6. Qiu2023, using the same SLAM method but with stellar parameter labels from APOGEE DR17, measured $T_\text{eff}$, log $g$, [M/H], and [$\alpha$/M] for 43,972 selected gM spectra in LAMOST DR9. Liang2022 adopted the PCA + LightGBM method, utilizing stellar parameter labels from APOGEE DR16 and photometric data, to determine $T_\text{eff}$, log $g$, and [M/H] for 37,979 stars.

Since none of the five studies provided quality flags for stellar parameters, and not all of them reported parameter errors, we uniformly applied the spectral quality criteria defined in Section~\ref{sec:external_comparison}, given that most studies derived stellar parameters from LAMOST spectra, to ensure the highest possible reliability of the parameters used in the comparison. Moreover, as different studies adopted varying methods to distinguish dM and gM stars, we aimed to minimize errors in stellar parameters caused by misclassification. To achieve this, when cross-matching with the five catalogs, we only included spectra for which the dM/gM classification was consistent across LAMOST, DGCM, and the Gaia CMD. Next, we will analyze the comparisons between the stellar parameters from this work and those provided by LASPM, Ding2022, Li2021, Qiu2023, and Liang2022 separately for dM and gM stars.

\begin{figure*}
	\centering
	\includegraphics[width=1\linewidth]{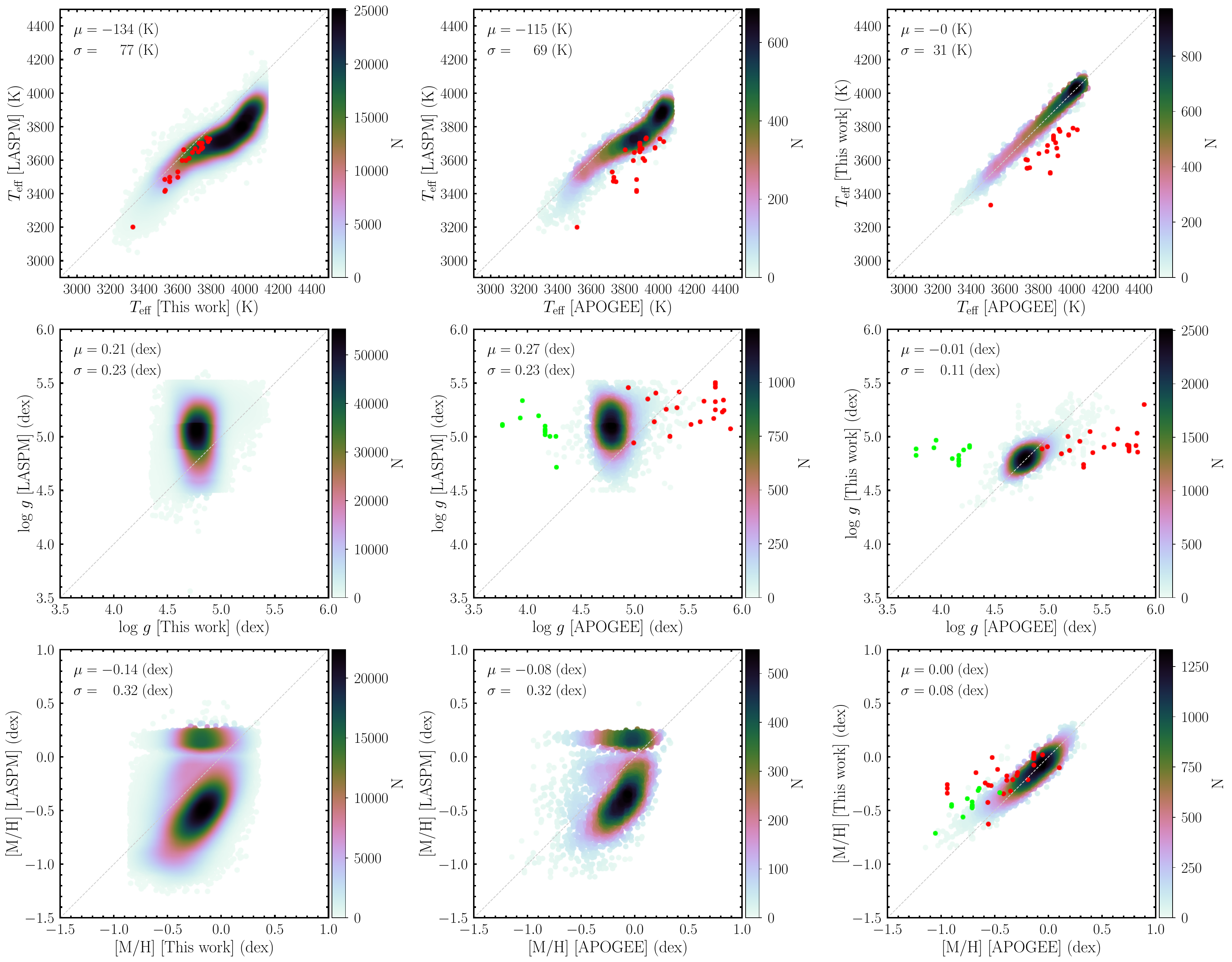}
	\caption{Comparison between the stellar parameters of dM stars derived in this work and those from the LAMOST Stellar Parameter Pipeline for M-type stars \citep[LASPM;][]{2021RAA....21..202D}. The left, middle, and right columns show parameter comparisons between LASPM and this work, LASPM and APOGEE, and this work and APOGEE, respectively. From top to bottom, the parameters are $T_\text{eff}$, log $g$, and [M/H]. Red dots indicate spectra for which $T_\text{eff}$ derived in this work and by APOGEE are inconsistent. Green dots indicate spectra with inconsistent log $g$ between this work and APOGEE. The red or green dots in Figures~\ref{fig:Ding2022_dM}, \ref{fig:Li2021_dM}, \ref{fig:kiel} and \ref{fig:teff_vs_bp_rp} have the same meaning as in this figure and will not be explained again. A detailed analysis of these points is provided in Sections \ref{sec:external_comparision_parameter} and \ref{sec:isochrone}. The $\mu$ and $\sigma$ in each panel represent the systematic difference and dispersion between the stellar parameters from different works, respectively.}
	\label{fig:LASPM_dM}
\end{figure*}

\begin{figure*}
	\centering
	\includegraphics[width=1\linewidth]{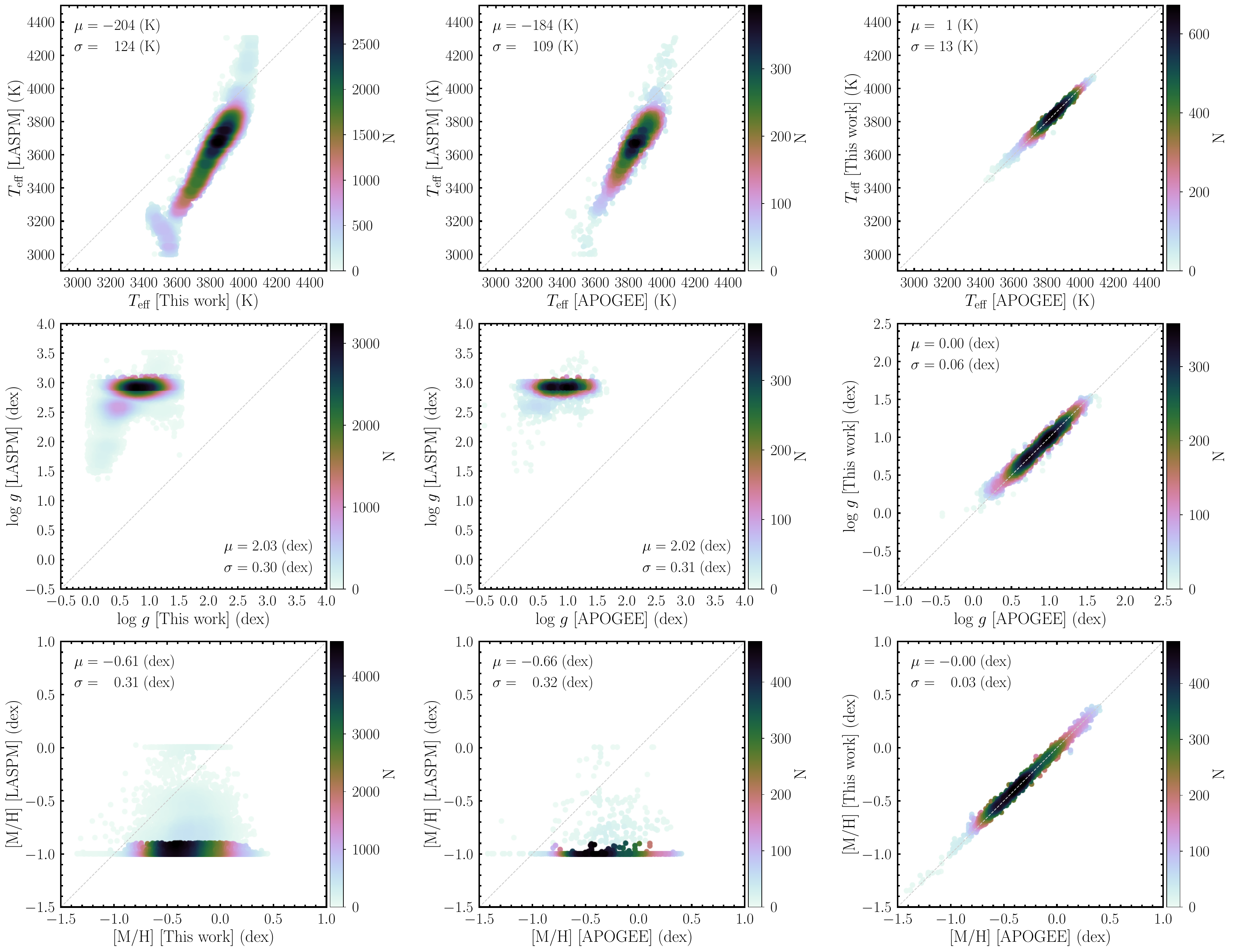}
	\caption{Comparison between the stellar parameters of gM stars derived in this work and those from LASPM. The left, middle, and right columns show comparisons between LASPM and this work, LASPM and APOGEE, and this work and APOGEE, respectively. From top to bottom, the parameters are $T_\text{eff}$, log $g$, and [M/H]. The $\mu$ and $\sigma$ in each panel represent the systematic difference and dispersion between the stellar parameters from different works, respectively.}
	\label{fig:LASPM_gM}
\end{figure*}

\begin{figure*}
	\centering
	\includegraphics[width=1\linewidth]{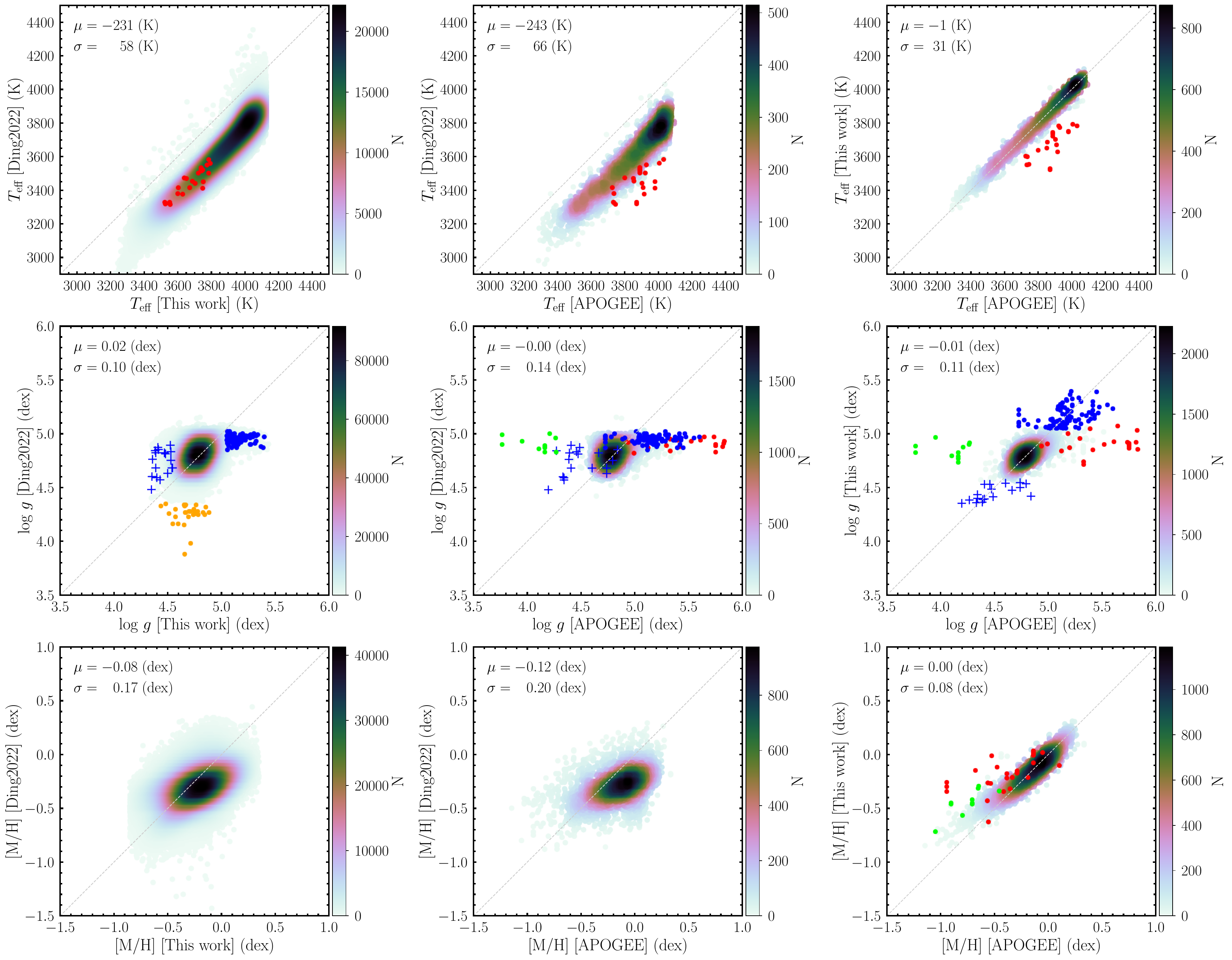}
	\caption{Comparison between the stellar parameters of dM stars derived in this work and those from \citet[hereafter Ding2022]{2022ApJS..260...45D}. The left, middle, and right columns show comparisons between Ding2022 and this work, Ding2022 and APOGEE, and this work and APOGEE, respectively. From top to bottom, the parameters are $T_\text{eff}$, log $g$, and [M/H]. The meanings of the red and green dots are the same as those explained in Figure~\ref{fig:LASPM_dM}. Blue dots and blue plus signs indicate spectra for which the log $g$ derived by Ding2022 and this work are inconsistent, in the regions where the log $g$ derived in this work is greater than 5.05 and less than 4.55, respectively. Orange dots indicate spectra with inconsistent log $g$ between Ding2022 and this work in the region where the log $g$ derived by Ding2022 is less than 4.35. The orange dots in Figure~\ref{fig:kiel} have the same meaning as in this figure and will not be explained again. A small number of dM spectra have log $g$ below 3.5 according to Ding2022, which are outside the range of the current figure. These spectra are marked as orange plus signs in Figure~\ref{fig:kiel}. A detailed discussion of these dots and plus signs is provided in Sections \ref{sec:external_comparision_parameter} and \ref{sec:isochrone}. The $\mu$ and $\sigma$ in each panel represent the systematic difference and dispersion between the stellar parameters from different works, respectively.}
	\label{fig:Ding2022_dM}
\end{figure*}

\begin{figure*}
	\centering
	\includegraphics[width=1\linewidth]{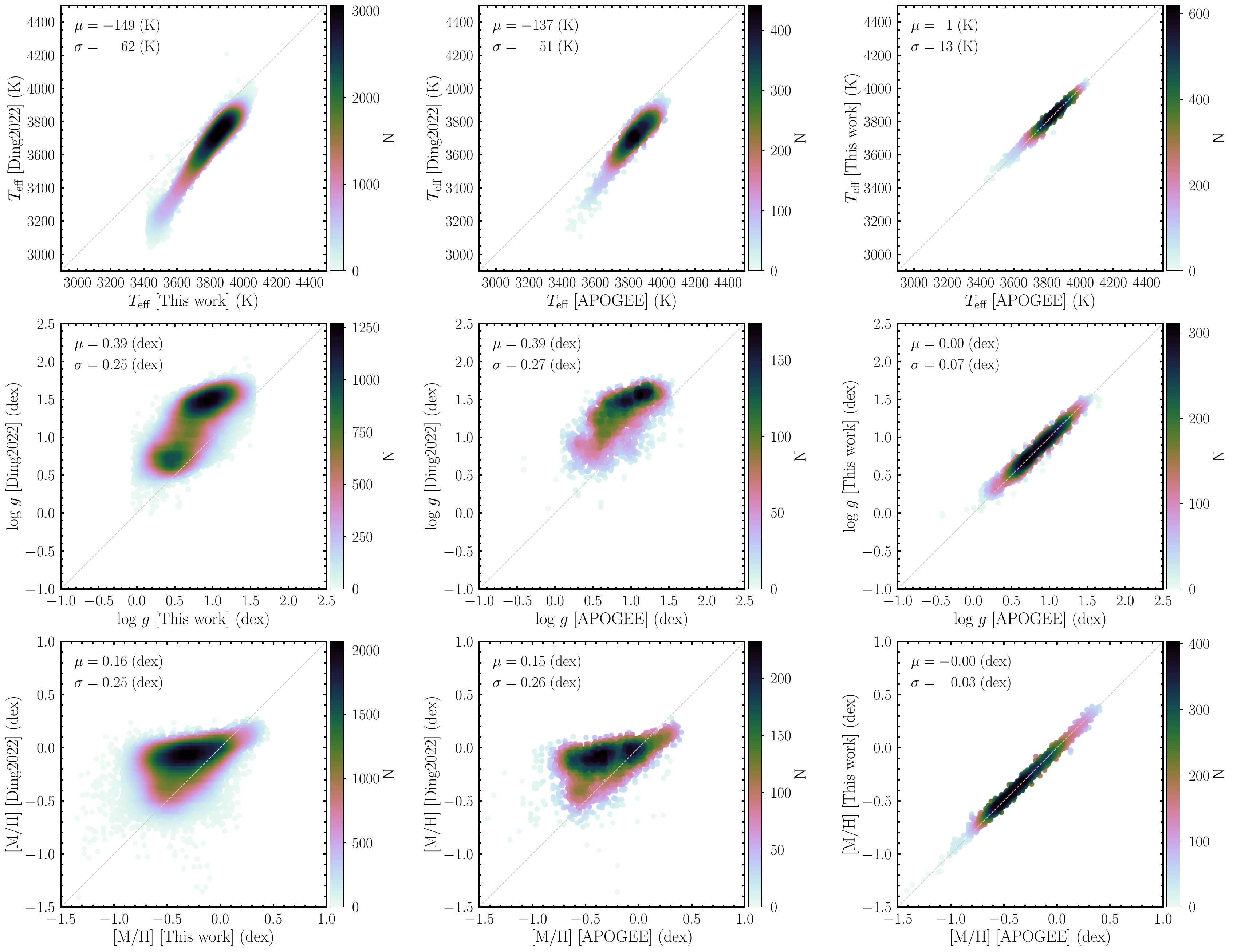}
	\caption{Comparison between the stellar parameters of gM stars derived in this work and those from Ding2022. The left, middle, and right columns show comparisons between Ding2022 and this work, Ding2022 and APOGEE, and this work and APOGEE, respectively. From top to bottom, the parameters are $T_\text{eff}$, log $g$, and [M/H]. The $\mu$ and $\sigma$ in each panel represent the systematic difference and dispersion between the stellar parameters from different works, respectively.}
	\label{fig:Ding2022_gM}
\end{figure*}

\begin{figure*}
	\centering
	\includegraphics[width=1\linewidth]{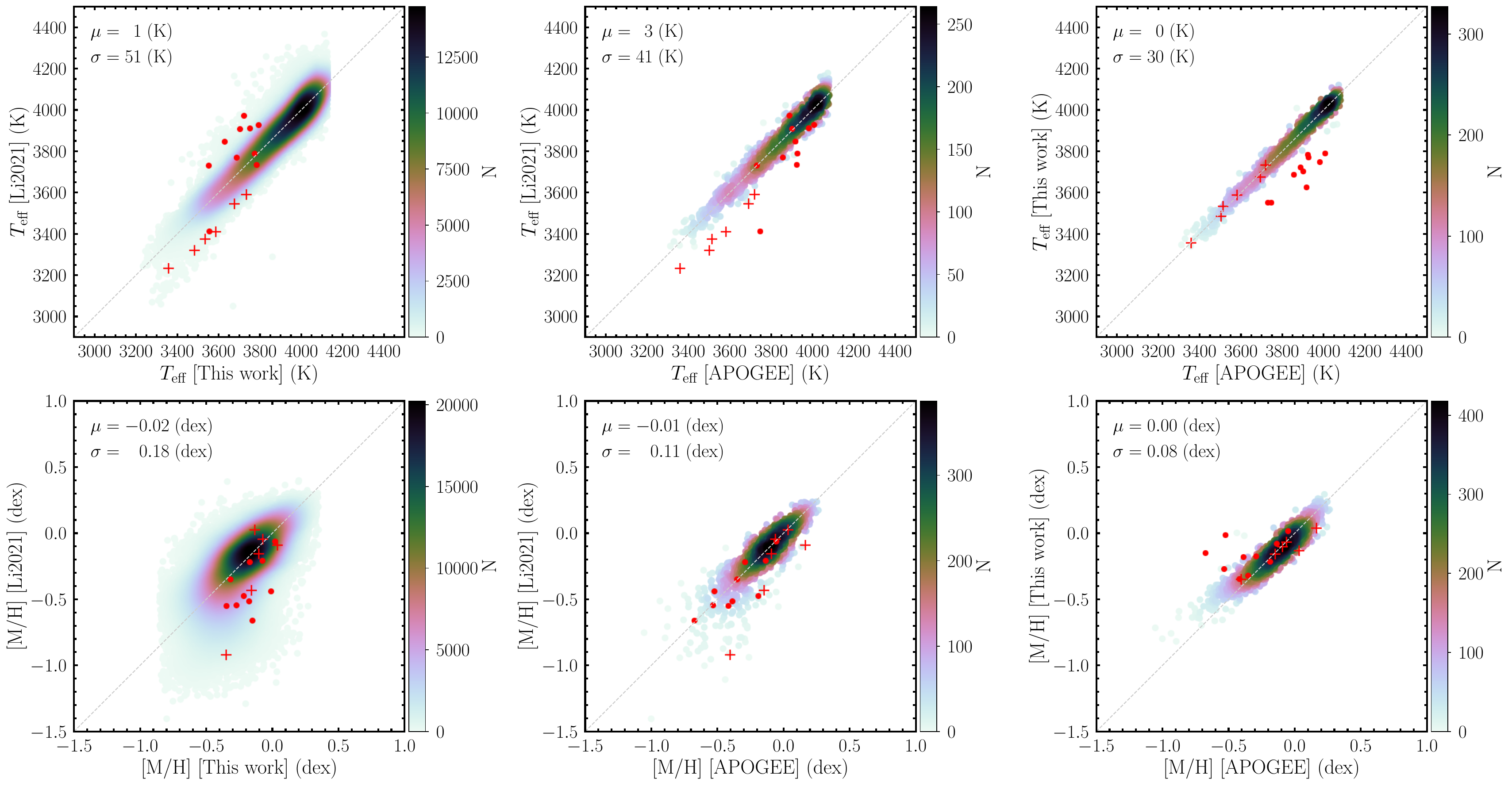}
	\caption{Comparison between the stellar parameters of dM stars derived in this work and those from \citet[hereafter Li2021]{2021ApJS..253...45L}. The left, middle, and right columns show comparisons between Li2021 and this work, Li2021 and APOGEE, and this work and APOGEE, respectively. From top to bottom, the parameters are $T_\text{eff}$ and [M/H]. The meanings of the red dots are the same as those explained in Figure~\ref{fig:LASPM_dM}. Red plus signs indicate spectra for which the $T_\text{eff}$ derived by Li2021 and APOGEE are inconsistent. The red plus signs in Figure~\ref{fig:teff_vs_bp_rp} have the same meaning and will not be explained again. A detailed discussion of these dots and plus signs is provided in Sections \ref{sec:external_comparision_parameter} and \ref{sec:isochrone}. The $\mu$ and $\sigma$ in each panel represent the systematic difference and dispersion between the stellar parameters from different works, respectively.}
	\label{fig:Li2021_dM}
\end{figure*}

\begin{enumerate}

\item LASPM. Figure~\ref{fig:LASPM_dM} and Figure~\ref{fig:LASPM_gM} display comparisons of $T_\text{eff}$, log $g$, and [M/H] for dM and gM stars between LASPM and this work in the left column. The middle and right columns show comparisons between LASPM and APOGEE, and this work and APOGEE, respectively. For dM stars, the systematic differences and dispersions between LASPM and this work are $-134\pm77$ K, $0.21\pm0.23$ dex, and $-0.14\pm0.32$ dex for $T_\text{eff}$, $\log g$, and [M/H], respectively. For gM stars, the corresponding values are $-204\pm124$ K, $2.03\pm0.30$ dex, and $-0.61\pm0.31$ dex. As shown in Figures~\ref{fig:LASPM_dM} and \ref{fig:LASPM_gM}, the stellar parameters measured in this work generally show better consistency with APOGEE. There exist systematic structures between LASPM and this work, which are also present in the comparison between LASPM and APOGEE, suggesting that these structures likely originate from LASPM. Possible reasons for these structures include: (1) The synthetic spectra used in LASPM and APOGEE are generated from different atmosphere models: BT-Settl and MARCS, respectively. (2) A significant discrepancy between the theoretical spectra used in LASPM and the observed spectra. (3) The parameter grid of the theoretical spectra is relatively coarse, with no interpolation performed within the grid. (4) No iterative optimization was applied to further reduce the differences between the theoretical and observed spectra when determining stellar parameters. For the stellar parameters of gM stars, the measurements in this work are highly consistent with those from APOGEE, suggesting that the discrepancies between LASPM and this work are most likely caused by LASPM. For dM stars, we conduct a detailed analysis of the discrepancies in Figure~\ref{fig:LASPM_dM}. The red points indicate spectra for which the $T_\text{eff}$ measured in this work is inconsistent with those from APOGEE. For these spectra, LASPM and APOGEE also show discrepancies in $T_\text{eff}$, whereas LASPM and this work are highly consistent. This suggests that APOGEE may have incorrectly measured $T_\text{eff}$ for these spectra. For most spectra, log $g$ measurements from LASPM show significant discrepancies with this work, similar to those between LASPM and APOGEE, indicating that the measurements in this work are possible more reliable. The outliers where log $g$ is overestimated in this work compared to APOGEE are marked with green points. For these green points, LASPM and APOGEE also show discrepancies in log $g$, suggesting that APOGEE may have incorrectly measured log $g$ for these spectra. For the outliers where log $g$ is underestimated in this work compared to APOGEE, we find that the vast majority correspond to the spectra (red points) for which APOGEE may have incorrectly measured $T_\text{eff}$. This suggests that the log $g$ measured by APOGEE for these spectra are also likely to be incorrect. Compared to APOGEE, the [M/H] provided in this work are systematically higher at the metal-poor end. This bias may result from the limited number of training samples in the metal-poor regime and potential labeling errors. It can be observed that most of the outliers with overestimated [M/H] at the metal-poor end correspond to the red and green points, which indicate spectra where APOGEE's parameter measurements may be problematic. In subsequent comparisons of stellar parameters with other studies, the causes of similar issues will not be discussed. Points and plus signs of the same color in later figures represent the same meanings. \label{item:LASPM}

\item Ding2022. Figures~\ref{fig:Ding2022_dM} and \ref{fig:Ding2022_gM} display comparisons of $T_\text{eff}$, log $g$, and [M/H] for dM and gM stars between Ding2022 and this work in the left column. The middle and right columns show comparisons between Ding2022 and APOGEE, and this work and APOGEE, respectively. For dM stars, the systematic differences and dispersions between Ding2022 and this work are $-231\pm58$ K, $0.02\pm0.10$ dex, and $-0.08\pm0.17$ dex for $T_\text{eff}$, log $g$, and [M/H], respectively. For gM stars, the corresponding values are $-149\pm62$ K, $0.39\pm0.25$ dex, and $0.16\pm0.25$ dex. Using the same analytical approach as in Comparison~\ref{item:LASPM}, the systematic structures between Ding2022 and this work, for both dM and gM stars, are likely caused by Ding2022. For dM stars, the analysis of the red and green points is consistent with that in Comparison~\ref{item:LASPM}. In the regions where log $g$ derived by this work is below 4.55 dex or above 5.05 dex, the measurements in this work show noticeable discrepancies with Ding2022. The spectra in these two regions that have APOGEE parameters are marked with blue plus signs and blue points, respectively. When comparing with APOGEE, the log $g$ values in this work exhibit larger dispersions but remain generally consistent, whereas the measurements from Ding2022 are mostly distributed in outlier positions, suggesting potential measurement issues in Ding2022. Additionally, Ding2022 includes a set of dM spectra with log $g$ values below 4.35 dex, which fall outside the reliable log $g$ range in this work. These spectra are marked with orange points. Since these spectra do not have APOGEE parameter cross-matches, they will be marked with the same color in the Kiel diagram (Figure~\ref{fig:kiel}) and analyzed accordingly. It is important to note that Ding2022 also contains another set of dM spectra with log $g$ values below 3.5 dex, which exceed the limits of Figure~\ref{fig:Ding2022_dM}, these spectra will be marked with orange plus signs in Figure~\ref{fig:kiel}. Both the orange points and orange plus signs will be analyzed in Section~\ref{sec:isochrone}.

\item Li2021. Figure~\ref{fig:Li2021_dM} displays comparisons of $T_\text{eff}$ and [M/H] for dM stars between Li2021 and this work in the left column. The middle and right columns show comparisons between Li2021 and APOGEE, and this work and APOGEE, respectively. The systematic differences and dispersions between Li2021 and this work are $1\pm51$ K and $-0.02\pm0.18$ dex for $T_\text{eff}$ and [M/H], respectively. Overall, the $T_\text{eff}$ and [M/H] from Li2021 and this work are consistent, as both adopted stellar parameter labels from APOGEE DR16. We find that compared to the $T_\text{eff}$ measured in this work, the $T_\text{eff}$ from Li2021 is systematically lower below 3600 K and systematically higher between 3600 K and 4000 K. The $T_\text{eff}$ from Li2021 also exhibit some systematically lower outliers below 3600 K when compared to APOGEE, these spectra are marked with red plus signs. It can be observed that both the red points and red plus signs are located in the region where Li2021 and this work show discrepancies in $T_\text{eff}$, whereas the red points lie in the region where Li2021 and APOGEE agree on $T_\text{eff}$. Based on the analysis in Comparison~\ref{item:LASPM}, the red points likely correspond to spectra for which APOGEE has incorrectly measured $T_\text{eff}$. We infer that the red plus signs may result from the SLAM-trained model in Li2021 being influenced by erroneous APOGEE $T_\text{eff}$ labels (red points), leading to systematic biases in the $T_\text{eff}$ from Li2021. This hypothesis will be further examined in the temperature-color diagrams (Figures~\ref{fig:teff_vs_bp_rp} and \ref{fig:teff_vs_w1_w2}). At the metal-poor end, the [M/H] values from Li2021 are systematically lower than those from APOGEE with a larger dispersion. This could be attributed to a combination of incorrect labels and an imbalanced distribution of training data.

\item Qiu2023. Figure~\ref{fig:Qiu2023_gM} displays comparisons of $T_\text{eff}$, log $g$, [M/H] and [$\alpha$/M] for gM stars between Qiu2023 and this work in the left column. The middle and right columns show comparisons between Qiu2023 and APOGEE, and this work and APOGEE, respectively. The systematic differences and dispersions between Qiu2023 and this work are $10\pm32$ K, $0.12\pm0.26$ dex, $-0.00\pm0.20$ dex and $-0.00\pm0.09$ dex, respectively. The $T_\text{eff}$ from Qiu2023 and this work are generally consistent. Through an analysis similar to that in Comparison~\ref{item:LASPM}, we find that the systematic structures in the other three parameters are likely caused by Qiu2023. These systematic structures may result from differences between the parameter labels in APOGEE DR16 and DR17, combined with the effects of different data-driven methods. Except for a small subset of spectra where APOGEE measures [$\alpha$/M] in the range $0.18<[\alpha/\text{M}]<0.24$ (marked with purple points), which are systematically underestimated in this work, the measurements from this work and APOGEE are largely consistent. This bias may be caused by an imbalance in the training data. For the purple points, Qiu2023 shows partial agreement with APOGEE, while the remaining cases exhibit an even more significant systematic underestimation.

\item Liang2022. Figures~\ref{fig:Liang2022_dM} and \ref{fig:Liang2022_gM} display comparisons of $T_\text{eff}$, log $g$, and [M/H] for dM and gM stars between Liang2022 and this work in the left column. The middle and right columns show comparisons between Liang2022 and APOGEE, and this work and APOGEE, respectively. For dM stars, the systematic differences and dispersions between Liang2022 and this work are $0\pm46$ K, $-0.04\pm0.16$ dex and $0.01\pm0.15$ dex for $T_\text{eff}$, log $g$, and [M/H], respectively. For gM stars, the corresponding values are $13\pm45$ K, $0.02\pm0.14$ dex and $0.01\pm0.15$ dex. For both dM and gM stars, the three parameters measured by Liang2022 are largely consistent with those from this work, as well as with APOGEE. For dM stars, the log $g$ from Liang2022 include some outliers below 4.5 dex, and its [M/H] exhibit systematic structures. Using a similar analytical approach as in Comparison~\ref{item:LASPM}, these issues are likely introduced by Liang2022. For gM stars, in comparison with APOGEE, the $T_\text{eff}$, log $g$, and [M/H] from Liang2022 show greater dispersion than those from this work, which may be due to the limited precision of the photometric data.

\end{enumerate}

Comparing each work to APOGEE, the stellar parameters provided in this work exhibit significant improvement in dispersion relative to the five other works. For dM stars, the dispersions in $T_\text{eff}$, log $g$, and [M/H] obtained by other works were 41--69 K, 0.14--0.23 dex, and 0.11--0.32 dex respectively, whereas this work achieves better dispersions of 31 K, 0.11 dex, and 0.08 dex. For gM stars, the dispersions in $T_\text{eff}$, log $g$, and [M/H] derived by other works were 32--109 K, 0.15--0.31 dex (exceeding 0.25 dex in all studies except Liang2022), and 0.15--0.32 dex, while this work achieves better dispersions of 13 K, 0.07 dex, and 0.03 dex. Additionally, Qiu2023 reported an [$\alpha$/M] dispersion of 0.07 dex, whereas this work achieves a dispersion of 0.02 dex. Overall, when comparing with APOGEE, the precision of parameters measured in this work improved by 21\%--75\% for dM stars and 53\%--91\% for gM stars, demonstrating that this approach substantially enhances the precision of parameter estimation for gM stars.

\begin{figure*}
	\centering
	\includegraphics[width=1\linewidth]{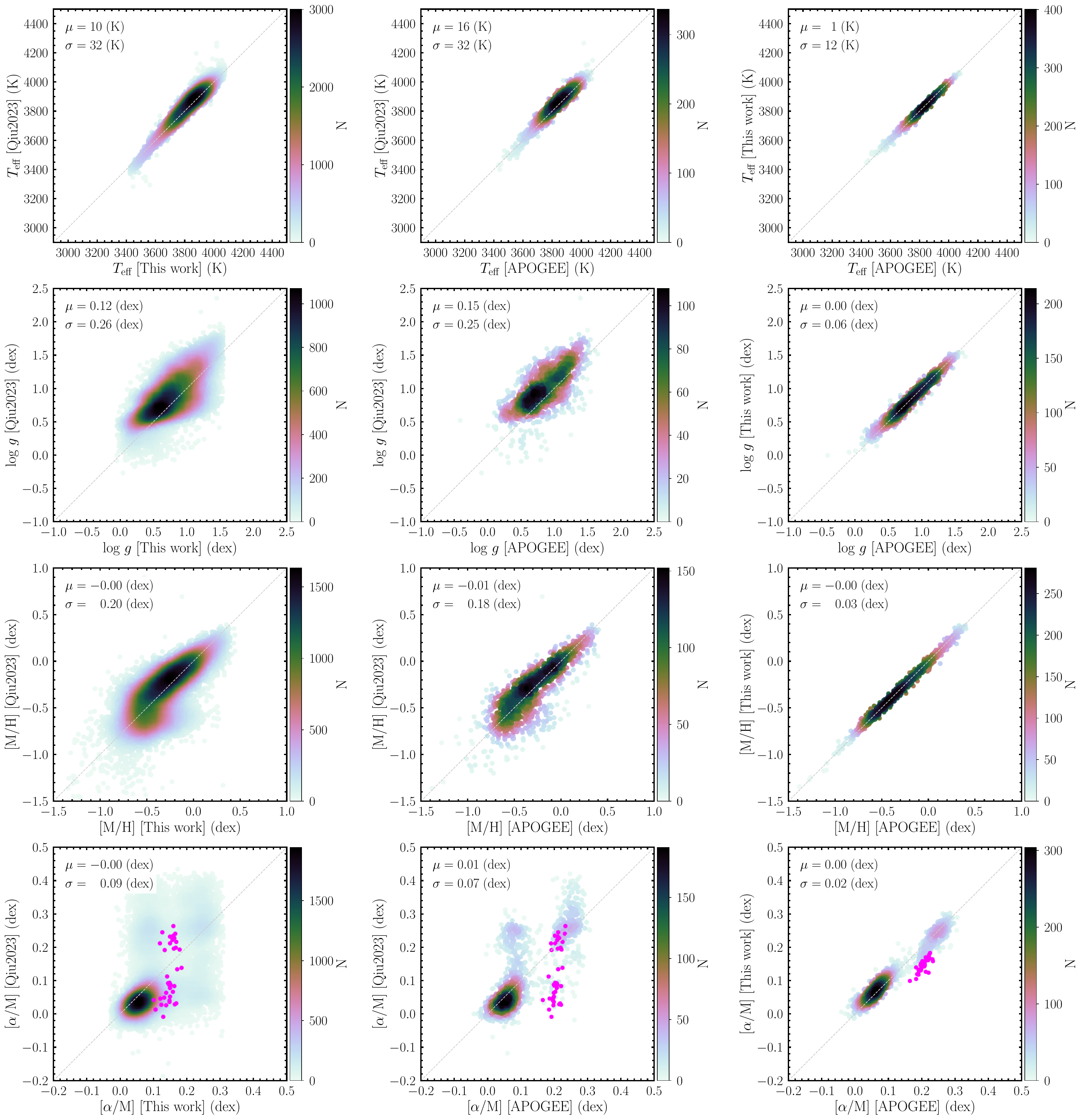}
	\caption{Comparison between the stellar parameters of gM stars derived in this work and those from \citet[hereafter Qiu2023]{2023RAA....23e5008Q}. The left, middle, and right columns show comparisons between Qiu2023 and this work, Qiu2023 and APOGEE, and this work and APOGEE, respectively. From top to bottom, the parameters are $T_\text{eff}$, log $g$, [M/H], and [$\alpha$/M]. Purple dots indicate spectra for which the [$\alpha$/M] derived in this work and by APOGEE are inconsistent. These points are discussed in Section~\ref{sec:external_comparision_parameter}. The $\mu$ and $\sigma$ in each panel represent the systematic difference and dispersion between the stellar parameters from different works, respectively.}
	\label{fig:Qiu2023_gM}
\end{figure*}

\begin{figure*}
	\centering
	\includegraphics[width=1\linewidth]{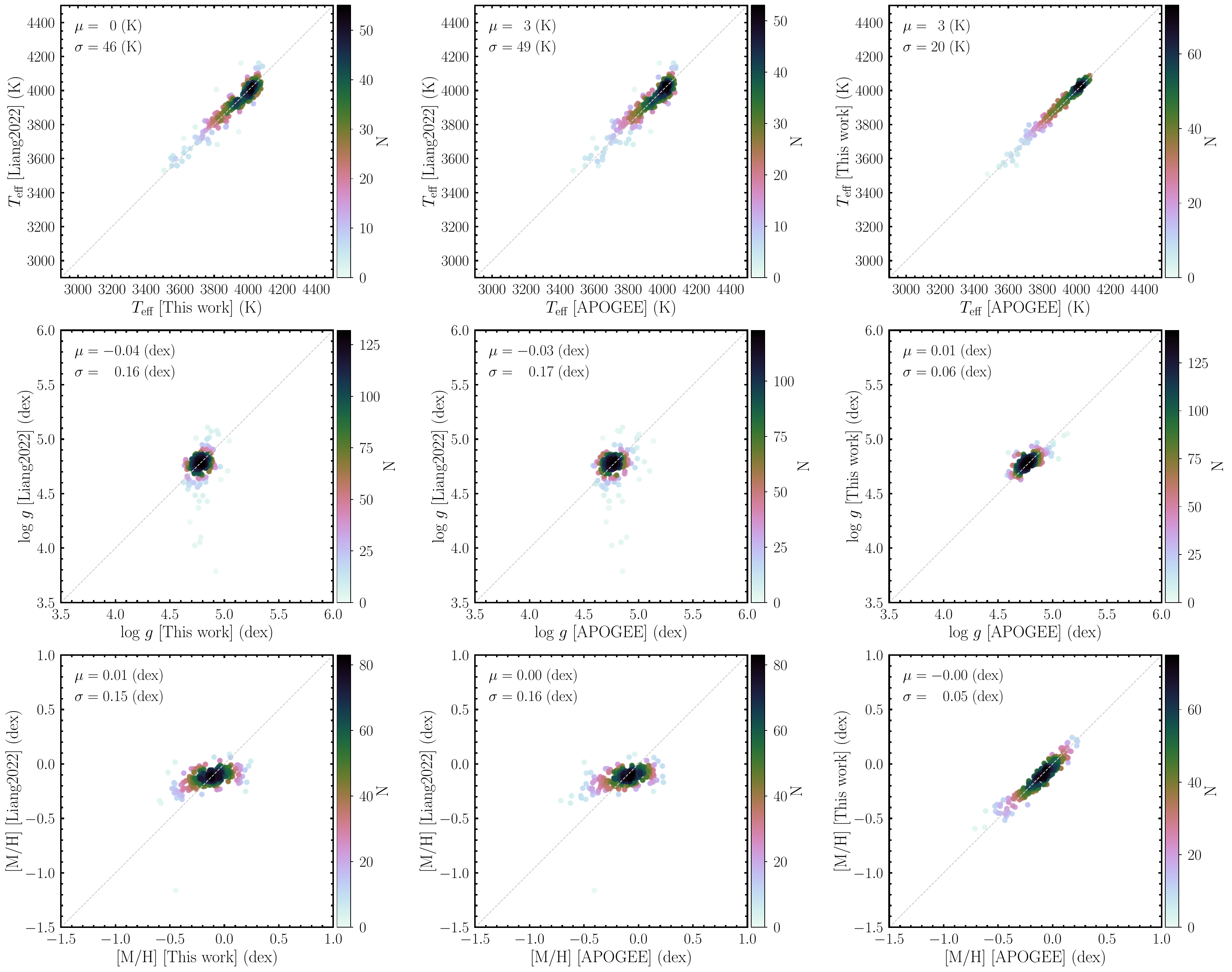}
	\caption{Comparison between the stellar parameters of dM stars derived in this work and those from \citet[hereafter Liang2022]{2022AJ....163..153L}. The left, middle, and right columns show comparisons between Liang2022 and this work, Liang2022 and APOGEE, and this work and APOGEE, respectively. From top to bottom, the parameters are $T_\text{eff}$, log $g$, and [M/H]. The $\mu$ and $\sigma$ in each panel represent the systematic difference and dispersion between the stellar parameters from different works, respectively.}
	\label{fig:Liang2022_dM}
\end{figure*}

\begin{figure*}
	\centering
	\includegraphics[width=1\linewidth]{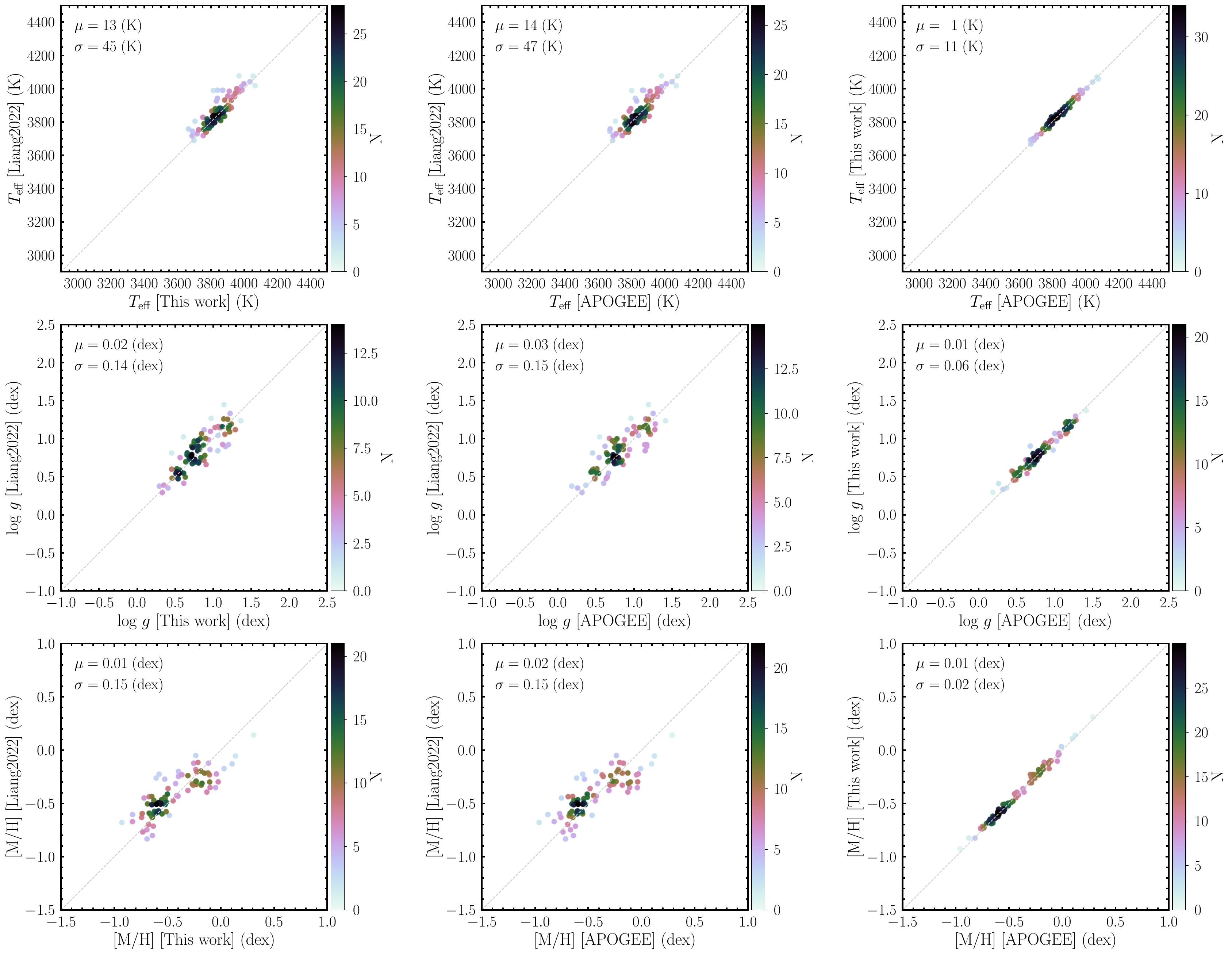}
	\caption{Comparison between the stellar parameters of gM stars derived in this work and those from Liang2022. The left, middle, and right columns show comparisons between Liang2022 and this work, Liang2022 and APOGEE, and this work and APOGEE, respectively. From top to bottom, the parameters are $T_\text{eff}$, log $g$, and [M/H]. The $\mu$ and $\sigma$ in each panel represent the systematic difference and dispersion between the stellar parameters from different works, respectively.}
	\label{fig:Liang2022_gM}
\end{figure*}

\subsection{Isochrone}
\label{sec:isochrone}

In Section~\ref{sec:external_comparision_parameter}, we analyzed the stellar parameters measured in this work, other studies, and APOGEE. The blue plus signs and points indicate spectra for which the log $g$ from LASPM and Ding2022 are inconsistent with those from this work, corresponding to regions where log $g$ in this work is either below 4.55 dex or above 5.05 dex. Previous analyses have demonstrated that the measurements in this work are more consistent with APOGEE, so no further analysis of the blue plus signs and points is necessary. The red points represent spectra for which the $T_\text{eff}$ in this work are inconsistent with those from APOGEE, while the green points indicate spectra with discrepancies in log $g$. Based on earlier analyses, we preliminarily conclude that the stellar parameters measured by APOGEE for these red and green points may be problematic. To ensure a more rigorous analysis, we further examine these points using the Kiel diagram. The orange points and plus signs indicate spectra for which the log $g$ measurements in Ding2022 and this work are inconsistent, corresponding to the regions where Ding2022 reported $3.5<$ log $g$ $<4.35$ dex and log $g$ $<3.5$ dex, respectively. Since these spectra do not have corresponding APOGEE parameters, they also require further investigation using the Kiel diagram.

\begin{figure*}
	\centering
	\includegraphics[width=0.95\linewidth]{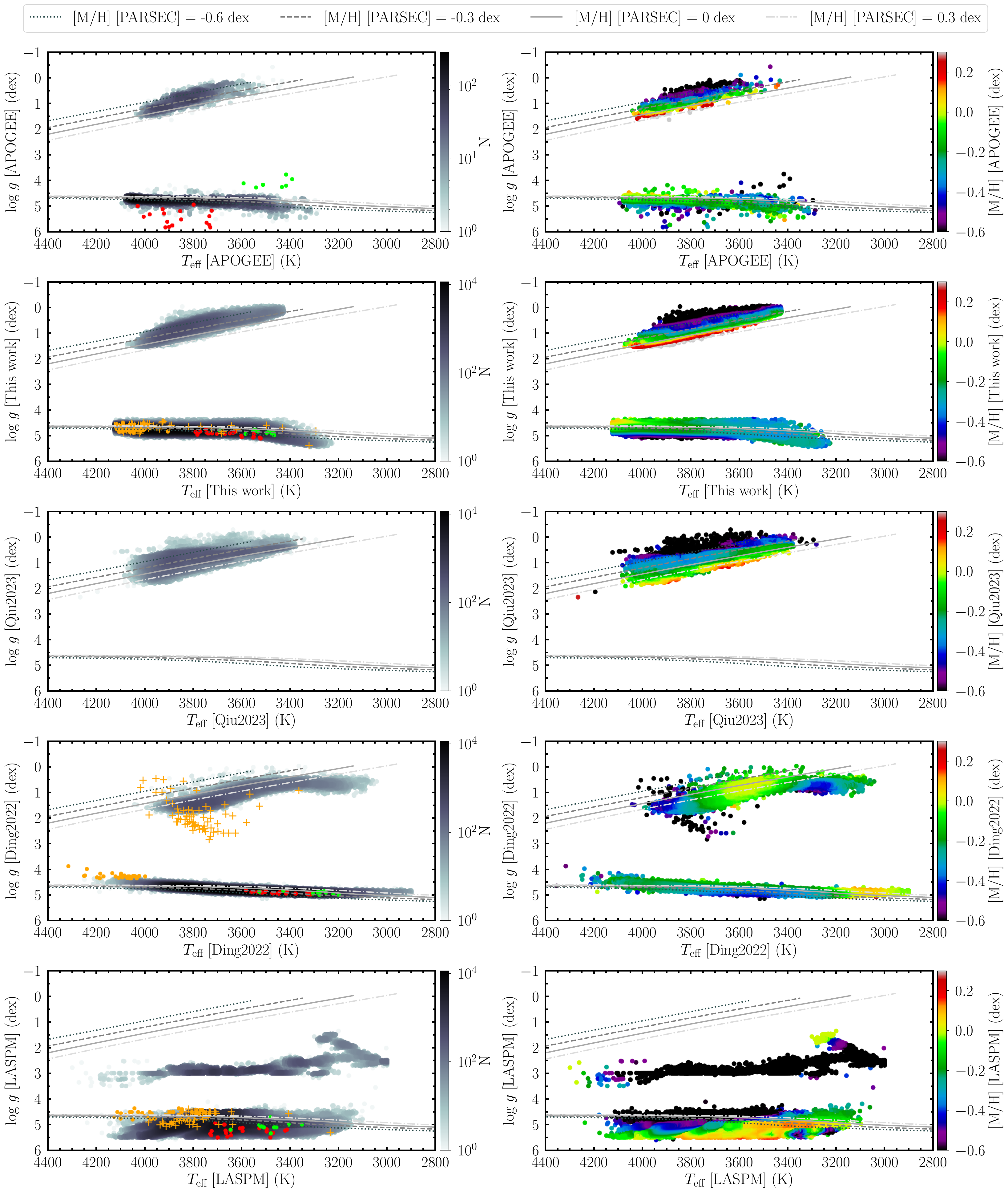}
	\caption{Kiel diagrams of this work and the comparison studies. From top to bottom, the panels show the Kiel diagrams from APOGEE, this work, Qiu2023, Ding2022, and LASPM. The meanings of the red and green dots in the left column are the same as those explained in Figure~\ref{fig:LASPM_dM}. The meanings of the orange dots and plus signs are the same as those explained in Figure~\ref{fig:Ding2022_dM}. A detailed discussion of these dots and plus signs is provided in Sections \ref{sec:external_comparision_parameter} and \ref{sec:isochrone}. The gray lines in each panel represent 9 Gyr theoretical isochrones generated using the PAdova and TRieste Stellar Evolution Code \citep[PARSEC;][]{2012MNRAS.427..127B}, with metallicities of –0.6, –0.3, 0.0, and 0.3 dex. The same PARSEC isochrone settings are adopted in Figures \ref{fig:teff_vs_bp_rp} and \ref{fig:teff_vs_w1_w2}, and will not be described again.}
	\label{fig:kiel}
\end{figure*}

Figure~\ref{fig:kiel} presents the Kiel diagrams of this work and the comparison studies. From top to bottom, the panels correspond to APOGEE, this work, Qiu2023, Ding2022, and LASPM. The gray lines represent 9 Gyr isochrones generated using the theoretical models from the PAdova and TRieste Stellar Evolution Code \citep[PARSEC;][]{2012MNRAS.427..127B}, with metallicities of -0.6, -0.3, 0, and 0.3 dex. It can be observed that the red and green points deviate from the isochrones in the Kiel diagram of APOGEE, whereas in the Kiel diagrams of this work, Ding2022, and LASPM, these points align more closely with the isochrones. This suggests that APOGEE likely encountered issues in the stellar parameter estimation for these red and green points. For the orange points, there is a trend of deviation from the isochrones in Ding2022's Kiel diagram, while in the Kiel diagrams of this work and LASPM, they align well with the isochrones, suggesting that Ding2022 likely made measurement errors. For the orange plus signs, they are located in the gM region in Ding2022's Kiel diagram, but in this work and LASPM's Kiel diagrams, they fall in the dM region and align with the isochrones. These spectra were manually checked and confirmed to be dMs, indicating that Ding2022 significantly misestimated the log $g$ for some dM spectra. It is worth noting that, similar to Figure~\ref{fig:LASPM_gM}, the $\log g$ values measured by LASPM for gM stars in Figure~\ref{fig:kiel} are concentrated around approximately 3, which differs significantly from the measurements obtained in this work and other studies. The possible reasons have been explained in Comparison~\ref{item:LASPM} of Section~\ref{sec:external_comparision_parameter}.

\begin{figure*}
	\centering
	\includegraphics[width=0.95\linewidth]{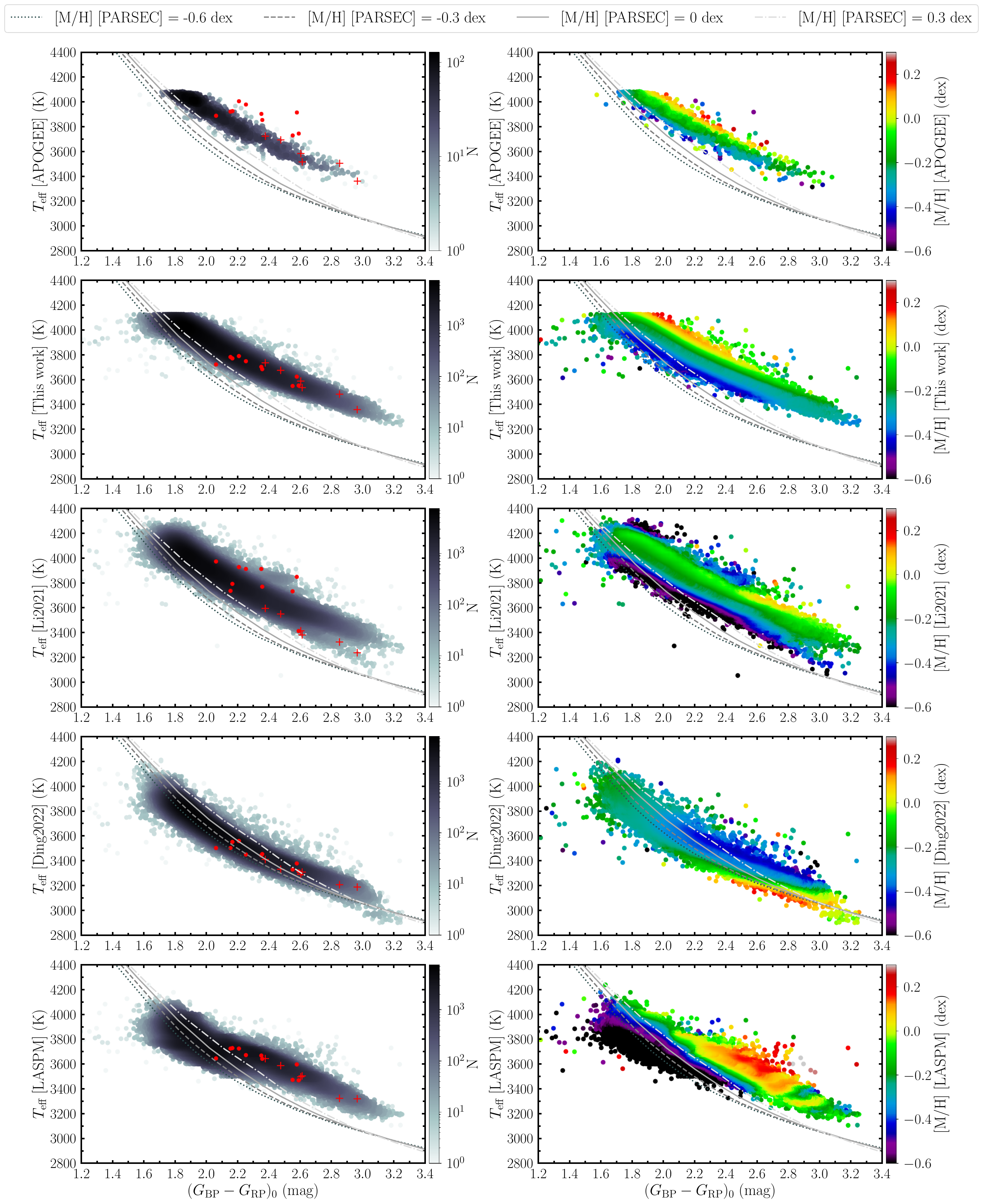}
	\caption{$T_\text{eff}$ vs $(G_\text{BP}-G_\text{RP})_0$ diagrams of dM stars from this work and the comparison studies. From top to bottom, the panels show the $T_\text{eff}$ vs $(G_\text{BP}-G_\text{RP})_0$ diagrams from APOGEE, this work, Li2021, Ding2022, and LASPM. The meanings of the red dots in the left column are the same as those explained in Figure~\ref{fig:LASPM_dM}. The meanings of the red plus signs are the same as those explained in Figure~\ref{fig:Li2021_dM}. A detailed discussion of these dots and plus signs is provided in Sections \ref{sec:external_comparision_parameter} and \ref{sec:isochrone}. The settings of the PARSEC isochrones (gray lines) shown in the panels are the same as those described in Figure~\ref{fig:kiel}.}
	\label{fig:teff_vs_bp_rp}
\end{figure*}

\begin{figure*}
	\centering
	\includegraphics[width=0.95\linewidth]{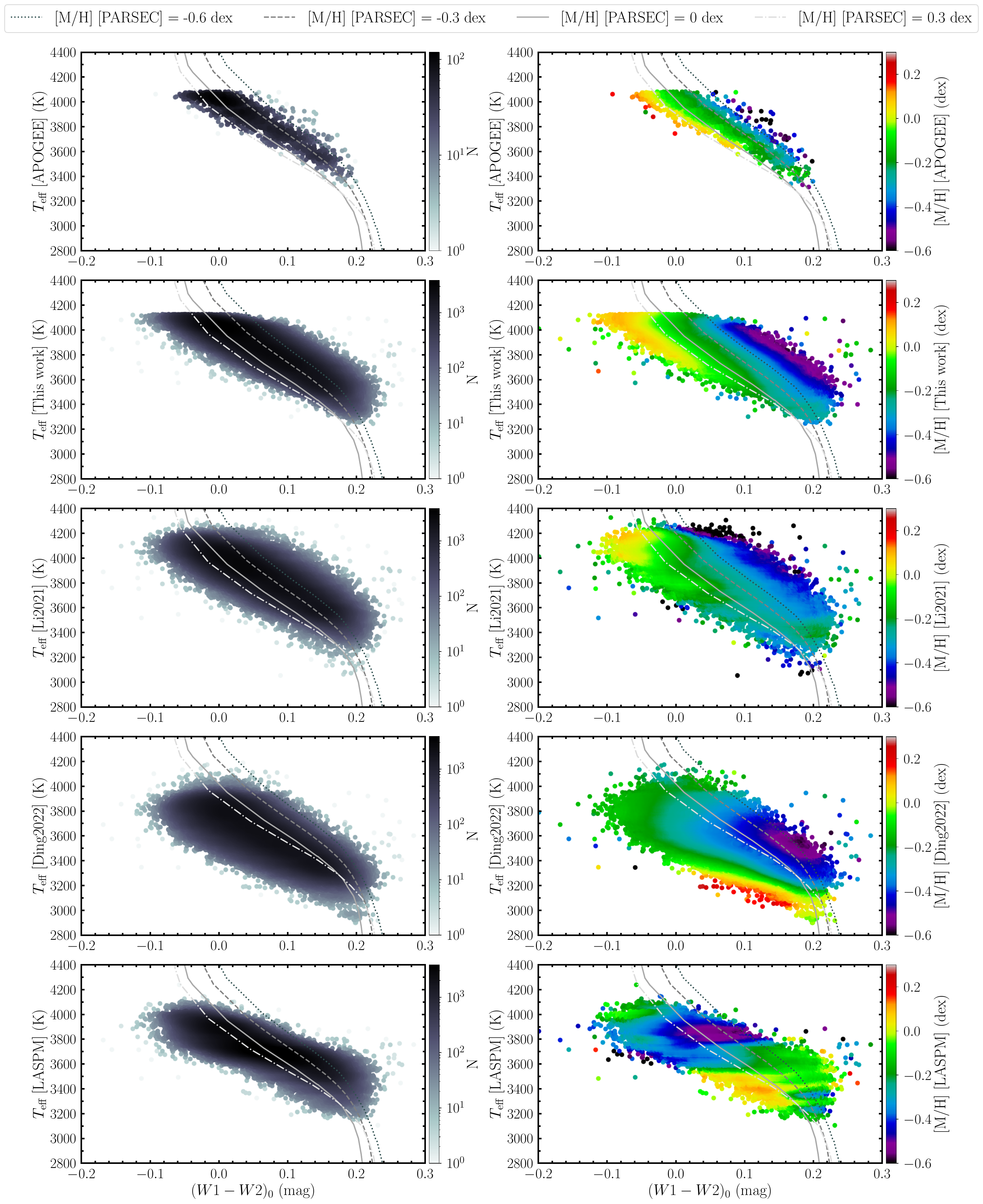}
	\caption{$T_\text{eff}$ vs $(W1-W2)_0$ diagrams of dM stars from this work and the comparison studies. From top to bottom, the panels show the $T_\text{eff}$ vs $(W1-W2)_0$ diagrams from APOGEE, this work, Li2021, Ding2022, and LASPM. The settings of the PARSEC isochrones (gray lines) shown in the panels are the same as those described in Figure~\ref{fig:kiel}.}
	\label{fig:teff_vs_w1_w2}
\end{figure*}

The Kiel diagram serves two purposes: (1) analyzing differences in stellar parameters between studies, and (2) evaluating metallicity gradients. It can be seen that, for both dM and gM stars, the metallicity gradients are more consistent between this work and APOGEE, following the trend of the isochrones. The metallicity gradient for dM stars is less pronounced, while for gM stars it is very clear. This may be due to the difficulty of achieving high-precision measurements within the narrow log $g$ range of dM stars. To analyze the metallicity gradient of dM stars without the effect of log $g$, we have plotted the $T_\text{eff}$ vs $(G_\text{BP}-G_\text{RP})_{0}$ diagram and the $T_\text{eff}$ vs $(W1-W2)_{0}$ diagram for this work and the comparison studies in Figures~\ref{fig:teff_vs_bp_rp} and \ref{fig:teff_vs_w1_w2}. From top to bottom, the panels correspond to APOGEE, this work, Li2021, Ding2022, and LASPM. The PARSEC isochrones in Figures~\ref{fig:teff_vs_bp_rp} and \ref{fig:teff_vs_w1_w2} are the same as those in Figure~\ref{fig:kiel}, and the extinction correction and quality selection for $(G_\text{BP}-G_\text{RP})_{0}$ are the same as those described in Section~\ref{sec:SCM}. The $W1$ and $W2$ magnitudes are from the AllWISE catalog \citep{2014yCat.2328....0C, 2012MNRAS.427..127B}, and extinction corrections were applied. The E($B-V$) values come from the 3D dust map Bayestar19 \citep{2019ApJ...887...93G} mentioned in Section~\ref{sec:SCM}, with extinction coefficients from \cite{2023ApJS..264...14Z}. To ensure the reliability of the $W1$ and $W2$ magnitudes, we only retained sources with $W1$ and $W2$ magnitude errors less than 0.03. From Figures~\ref{fig:teff_vs_bp_rp} and \ref{fig:teff_vs_w1_w2}, it can be seen that for the metallicity gradient of dM stars, this work is more consistent with APOGEE and shows a clearer trend, following the variation of the isochrones.

In the analysis in Section~\ref{sec:external_comparision_parameter}, the red plus signs indicate spectra for which the $T_\text{eff}$ from Li2021 and APOGEE are inconsistent. It is preliminarily believed that the SLAM model trained by Li2021 was influenced by the incorrect labels from APOGEE (red points). Since Li2021 did not measure log $g$, this work can only use the $T_\text{eff}$ vs $(G_\text{BP}-G_\text{RP})_{0}$ diagram (Figure~\ref{fig:teff_vs_bp_rp}) for further analysis. From Figure~\ref{fig:teff_vs_bp_rp}, it can be seen that the red points and plus signs are located in the anomalous region of Li2021's $T_\text{eff}$ vs $(G_\text{BP}-G_\text{RP})_{0}$ diagram, while they are in the normal region in the $T_\text{eff}$ vs $(G_\text{BP}-G_\text{RP})_{0}$ diagrams of this work, Ding2022, and LASPM. In APOGEE's $T_\text{eff}$ vs $(G_\text{BP}-G_\text{RP})_{0}$ diagram, the red points are in the anomalous region, while the red plus signs are in the normal region. This further confirms that APOGEE made errors in the $T_\text{eff}$ for the red points, which in turn affected the SLAM model trained by Li2021, leading to incorrect $T_\text{eff}$ for the red points and plus signs in Li2021's results. 

\section{Summary}
\label{sec:Summary}

This work first designed and successfully implemented a fast and efficient process tailored to the characteristics of the vast amount of low-resolution LAMOST spectral data (see Figure~\ref{workflow_diagram}). This process corrected the spectral types, dM and gM classifications, and RVs of the LAMOST DR10 M-type star catalog, ensuring the reliability of the corrections through manual verification, thereby constructing a cleaner M-type star sample. Secondly, we trained ten CNN models for dM and gM spectra using ten-fold cross-validation and model ensemble methods, predicting $T_\text{eff}$, log $g$, [M/H], [$\alpha$/M], and their associated errors. Finally, we conducted quality assessment cuts on the RVs and stellar parameters provided in this work and presented a Recommended Catalog (see Table~\ref{tab:recommended_catalog}). External comparisons confirm that the RVs and stellar parameters recommended in this work exhibit higher reliability. The main contributions of this work can be summarized as follows:

\begin{enumerate}
\item Identification and validation of non-M spectra. We trained two CNN classification models using LAMOST spectra and more accurate filtered classification labels (an eight-class model: OB, A, F, G, K, M, GALAXY and QSO, and a two-class model: WD and M) to classify all spectra in the LAMOST M-type star catalog, and by combining the positions of M-type stars in the Gaia CMD, we identified 22,496 non-M spectra after manual verification, of which 16,880 spectra have been updated in LAMOST DR8 and subsequent data releases.

\item Correction of dM and gM classifications. We trained a CNN model for dM/gM classification using LAMOST spectra and more accurate filtered classification labels to classify all M-type star spectra, and by combining the positions of dM and gM in the Gaia CMD and the dM/gM classification results provided by LAMOST, manually corrected the dM/gM classification for 2,078 spectra.

\item Correction of RVs for M-type stars. We re-measured the RVs of all M-type spectra using the template matching and Doppler shift methods, calculating the corresponding errors. By comparing with the RVs from LAMOST, we manually checked the spectra with potentially incorrect LAMOST RVs and those with LAMOST $|\text{RV}|>150\ \text{km s}^{-1}$, and corrected the RVs for 12,900 spectra. It is important to note that we have flagged 51,668 spectra with potential issues in the wavelength calibration lamps before January 14, 2012, as well as 89 spectra for which the RVs could not be determined due to spectral quality issues in the Recommended Catalog. The RVs of these spectra may be problematic and should be used with caution. The dispersion between the RVs recommended in this work and Gaia DR3 is 11 km~s$^{-1}$, which reduces to 7 km~s$^{-1}$ when the S/N is greater than 20. The systematic difference between the two is approximately -5 km~s$^{-1}$, which is likely due to the wavelength calibration lamps used by LAMOST.

\item Measurement of M-type stellar parameters. We used a ten-fold cross-validation approach to train ten CNN regression models for dM and gM spectra, with stellar parameter labels from APOGEE DR16. The mean and standard deviation of the predictions from the ten models were taken as the final stellar parameters and their associated errors. Using this model ensemble method, we predicted $T_\text{eff}$, log $g$, [M/H], [$\alpha$/M], and their errors for 820,493 dM spectra and 50,025 gM spectra. It is important to note that we performed rigorous quality assessment cuts of the stellar parameters (see Section~\ref{sec:quality_assessment}) and flagged them in the recommended catalog. The number of recommended dM parameters is 716,207 (87\% of the total dM spectra), and the number of recommended gM parameters is 40,907 (82\% of the total gM spectra). For dM stars, the recommended stellar parameter ranges are: $3225\ \text{K} < T_\text{eff} < 4125\ \text{K}$, $4.35\ \text{dex} < \text{log}\ g < 5.45\ \text{dex}$, $-0.85\ \text{dex} < [\text{M}/\text{H}] < 0.35\ \text{dex}$ and $-0.13\ \text{dex} < [\alpha/\text{M}] < 0.25\ \text{dex}$; for gM stars, they are: $3425\ \text{K} < T_\text{eff} < 4075\ \text{K}$, $-0.05\ \text{dex} < \text{log}\ g < 1.55\ \text{dex}$, $-1.35\ \text{dex} < [\text{M}/\text{H}] < 0.45\ \text{dex}$ and $-0.03\ \text{dex} < [\alpha/\text{M}] < 0.31\ \text{dex}$. For dM stars, the average internal errors for $T_\text{eff}$, log $g$, [M/H], and [$\alpha$/M] are 30 K, 0.07 dex, 0.07 dex and 0.02 dex, respectively; for gM stars, they are 17 K, 0.07 dex, 0.05 dex and 0.02 dex. For dM stars, the errors between this work and APOGEE DR16 are: $-0\pm34$ K, $0.00\pm0.12$ dex, $0.00\pm0.09$ dex and $0.00\pm0.03$ dex; for gM stars, they are: $1\pm14$ K, $0.00\pm0.07$ dex, $0.00\pm0.04$ dex and $0.00\pm0.02$ dex.

\item Comparison of stellar parameters with other studies. Firstly, this work used APOGEE results as a reference benchmark to evaluate the results of this work and five other M-type stellar parameter measurement studies. For dM stars, the precision of $T_\text{eff}$, log $g$, and [M/H] measurements in this work shows improvements of 24\%--55\%, 21\%--52\%, and 27\%--75\% respectively compared to the five other studies; for gM stars, the precision of $T_\text{eff}$, log $g$, [M/H], and [$\alpha$/M] improves by 59\%--88\%, 53\%--77\%, 80\%--91\%, and 71\% respectively. Compared to five previous studies, the parameters derived from this work demonstrate greater consistency with APOGEE while simultaneously achieving a remarkably significant improvement in parameter measurement precision for gM stars. Secondly, In the Kiel diagram, $T_\text{eff}$ vs $(G_\text{BP}-G_\text{RP})_{0}$ diagram, and $T_\text{eff}$ vs $(W1-W2)_{0}$ diagram, the metallicity gradient of this work aligns more closely with APOGEE than other five studies. 

\end{enumerate}

The advantages of the stellar parameters provided in this work are as follows: (1) The ``label transfer + parameter prediction" strategy applied in this work, which used LAMOST spectra and high-precision APOGEE parameters as training samples, effectively reduces systematic errors arising from spectral mismatches compared to directly applying synthetic-trained models to observed spectra; compared to parameter inference methods, the ``parameter prediction" method can provide more reliable parameter measurements with greater efficiency. (2) Compared to traditional machine learning algorithms, CNNs have superior feature extraction and nonlinear modeling capabilities, can effectively process high-dimensional data and capture correlations between different parameters. (3) The ten-fold cross-validated CNN ensemble architecture effectively reduces dependence on an individual dataset and minimizes overfitting risks when training data is limited, enhancing both the reliability and stability of stellar parameter determinations.

The limitations of the stellar parameters provided in this work, using APOGEE as the reference benchmark, are as follows: (1) dM spectra in the range $-1.1\ \text{dex} < [\text{M}/\text{H}] < -0.5\ \text{dex}$ are systematically overestimated by approximately 0.1 dex. (2) dM spectra in the range $[\alpha/\text{M}] > 0.14\ \text{dex}$ are systematically underestimated by approximately 0.1 dex. (3) gM spectra in the range $0.18\ \text{dex} < [\alpha/\text{M}] < 0.24\ \text{dex}$ are systematically underestimated by approximately 0.08 dex. These limitations are likely due to an imbalance in the training data in the above-mentioned parameter ranges. In the future, we will attempt to address the data imbalance issue using the following approaches: (1) Weight adjustment. During training, we can adjust the sample weights in the loss function to make the model more sensitive to less frequently occurring stellar parameter ranges. Weight adjustment helps the model focus more on regions with fewer samples, reducing bias. (2) Data augmentation. For relatively scarce stellar parameter samples, we can use specific data augmentation techniques, such as adding noise, oversampling, or using Generative Adversarial Networks (GANs) to generate missing samples, thus balancing the data and increasing its diversity.

Beyond the issues in stellar parameter measurements, the Recommended Catalog exhibits inherent constraints. (1) Despite implementing multiple criteria and conducting rigorous visual inspections of over 100,000 spectra to minimize contamination from non-M-type stars, the Recommended Catalog may still contain a small number of non-M spectra, mainly including low-quality UNKNOWN spectra and late K-type stars\footnote{Given the extreme similarity between the spectral features of late K-type and early M-type stars, reliable discrimination by both algorithm and visual inspection are difficult. Therefore, we roughly estimated the possible fraction of UNKNOWN spectra in the Recommended Catalog, which is approximately 0.8\% (computed by 419/50,000; the two value were mentioned in Section~\ref{sec:DGCM}), and based on our experience, we consider the fraction of late K-type contaminants to be roughly comparable in magnitude.}, for which the DGCM, RVMM, and SPMM analyses may exhibit relatively larger uncertainties. (2) Since the Recommended Catalog is constructed based on the official LAMOST M-type star catalog, it is acknowledged that genuine M-type stars misclassified by the LAMOST pipeline as other spectral types may have been inadvertently excluded. Empirically, the misclassification rate of the LAMOST pipeline is estimated to be approximately 3\%, which suggests that the Recommended Catalog encompasses the vast majority of M-type stars observed by LAMOST. Nevertheless, to ensure the construction of a more comprehensive and complete M-type star catalog, future systematic searches through the entire LAMOST dataset will be necessary to identify and recover these potentially overlooked M-type candidates. (3) In constructing the Recommended Catalog, we excluded only the obvious spectral binaries identified during visual inspection within the SCM module, while the small fraction of spectral binaries present in the non-visually inspected spectra remained unfiltered. Additionally, we did not employ other criteria to specifically identify or exclude other types of binary systems. Although known binary fraction for M dwarfs is probably relatively lower \citep{2019AJ....157..216W}, the current parameter estimation methods may exhibit reduced accuracy when applied to binary spectra. To minimize the impact of binaries on parameter estimates, users can apply the \texttt{CMD\_flag} = 1 filter in the Recommended Catalog to exclude potential binaries and sources with unreliable Gaia parameters.

In addition to the low-resolution observations, LAMOST medium-resolution survey has released $\sim$~10 million coadded spectra, with the official LAMOST catalog providing stellar parameters for only about one-quarter of the spectra. In the future, we will attempt to apply the methods proposed in this work to the stellar parameter measurements of M-type stars in LAMOST medium-resolution spectra.

\section*{Acknowledgements}

We thank Zipeng Zheng, Rui Wang, Jingyi Zhang, Yanxin Guo, Bing Du, Xiao Kong, Fang Zuo, Kefei Wu, Wen Hou, Yihan Song, Xiaoting Fu, Mingyi Ding, Jiadong Li, Bo Zhang, Peng Wei, Shuguo Ma, Yuanhao Wen, Xianglei Chen and Caixia Qu for useful discussions. This work is supported by the National Natural Science Foundation of China (12273078, 12411530071, 12273075) and the National Astronomical Observatories of the Chinese Academy of Sciences (No.E4ZR0516). We also acknowledge support from Royal Society IEC\textbackslash NSFC\textbackslash 233140 exchange grant. Guoshoujing Telescope (the Large Sky Area Multi-Object Fiber Spectroscopic Telescope, LAMOST) is a National Major Scientific Project built by the Chinese Academy of Sciences. Funding for the Project has been provided by the National Development and Reform Commission. LAMOST is operated and managed by the National Astronomical Observatories, Chinese Academy of Sciences. This research makes use of data from the European Space Agency (ESA) mission Gaia, processed by the Gaia Data Processing and Analysis Consortium. Funding for the Sloan Digital Sky Survey IV has been provided by the Alfred P. Sloan Foundation, the U.S. Department of Energy Office of Science, and the Participating Institutions. SDSS acknowledges support and resources from the Center for High-Performance Computing at the University of Utah. The SDSS website is \url{www.sdss.org}. This publication makes use of data products from the Wide-field Infrared Survey Explorer, which is a joint project of the University of California, Los Angeles, and the Jet Propulsion Laboratory/California Institute of Technology, funded by the National Aeronautics and Space Administration.

This research also makes use of Astropy, a community-developed core Python package for Astronomy \citep{2013A&A...558A..33A}, the TOPCAT tool \citep{2005ASPC..347...29T} and the VizieR catalog access tool and the Simbad database, operated at Centre de Donnees astronomiques de Strabourg (CDS), France.

$Facilities$: LAMOST, Gaia, SDSS, WISE.

\bibliography{M_DR10}{}
\bibliographystyle{aasjournal}
\end{CJK*}
\end{document}